\pdfoutput=1
\documentclass[%
prx,twocolumn,
notitlepage,
superscriptaddress,
 amsmath,amssymb,
 aps,
prx,
floatfix,
longbibliography
]{revtex4-1}
\usepackage[T1]{fontenc}

\usepackage{color}
\usepackage{amsmath}
\usepackage{amssymb}
\usepackage{bm}
\usepackage{graphicx}
\usepackage[usenames,dvipsnames]{xcolor}
\usepackage[%
  colorlinks=true,
  urlcolor=blue,
  linkcolor=blue,
  citecolor=blue
]{hyperref}

\usepackage{array}
\definecolor{bleu}{rgb}{0.16,0.2.5,0.36}
\definecolor{darkGreen}{RGB}{4,161,85}

\newcommand{\EDF}{Extended Data Fig.}

\begin{document}
\pdfoutput=1

\title{Auxetic Granular Metamaterials}
%
\author{Daan Haver}
\thanks{These two authors contributed equally}
\affiliation{Institute of Physics, Universiteit van Amsterdam, 1098 XH Amsterdam, The Netherlands}
\author{Daniel Acu\~{n}a}
\thanks{These two authors contributed equally}
\affiliation{Instituto de Física, Pontificia Universidad Católica de Chile, 8331150 Santiago, Chile}
\author{Shahram Janbaz}
\affiliation{Institute of Physics, Universiteit van Amsterdam, 1098 XH Amsterdam, The Netherlands}
\author{Edan Lerner}
\affiliation{Institute of Physics, Universiteit van Amsterdam, 1098 XH Amsterdam, The Netherlands}
\author{Gustavo D\"{u}ring}
\affiliation{Instituto de Física, Pontificia Universidad Católica de Chile, 8331150 Santiago, Chile}
\author{Corentin Coulais}
\affiliation{Institute of Physics, Universiteit van Amsterdam, 1098 XH Amsterdam, The Netherlands}

\date{October 5, 2023}
\maketitle

{\bf 
The flowing, jamming and avalanche behavior of granular materials is satisfyingly universal and vexingly hard to tune: a granular flow is typically intermittent and will irremediably jam if too confined. 
Here, we show that granular metamaterials made from particles with a negative Poisson's ratio yield more easily and flow more smoothly than ordinary granular materials. 
We first create a collection of auxetic grains based on a re-entrant mechanism and show that each grain exhibits a negative Poisson's ratio regardless of the direction of compression.
Interestingly, we find that the elastic and yielding properties are governed by the high compressibility of granular metamaterials: at a given confinement they exhibit lower shear modulus, lower yield stress and more frequent, smaller avalanches than materials made from ordinary grains. 
We further demonstrate that granular metamaterials promote flow in more complex confined geometries, such as intruder and hopper geometries, even when the packing contains only a fraction of auxetic grains. 
Our findings blur the boundary between complex fluids and metamaterials and could help in  scenarios that involve process, transport and reconfiguration of granular materials.}

\begin{figure}[b!]
\centering
\includegraphics[width=1.0\linewidth]{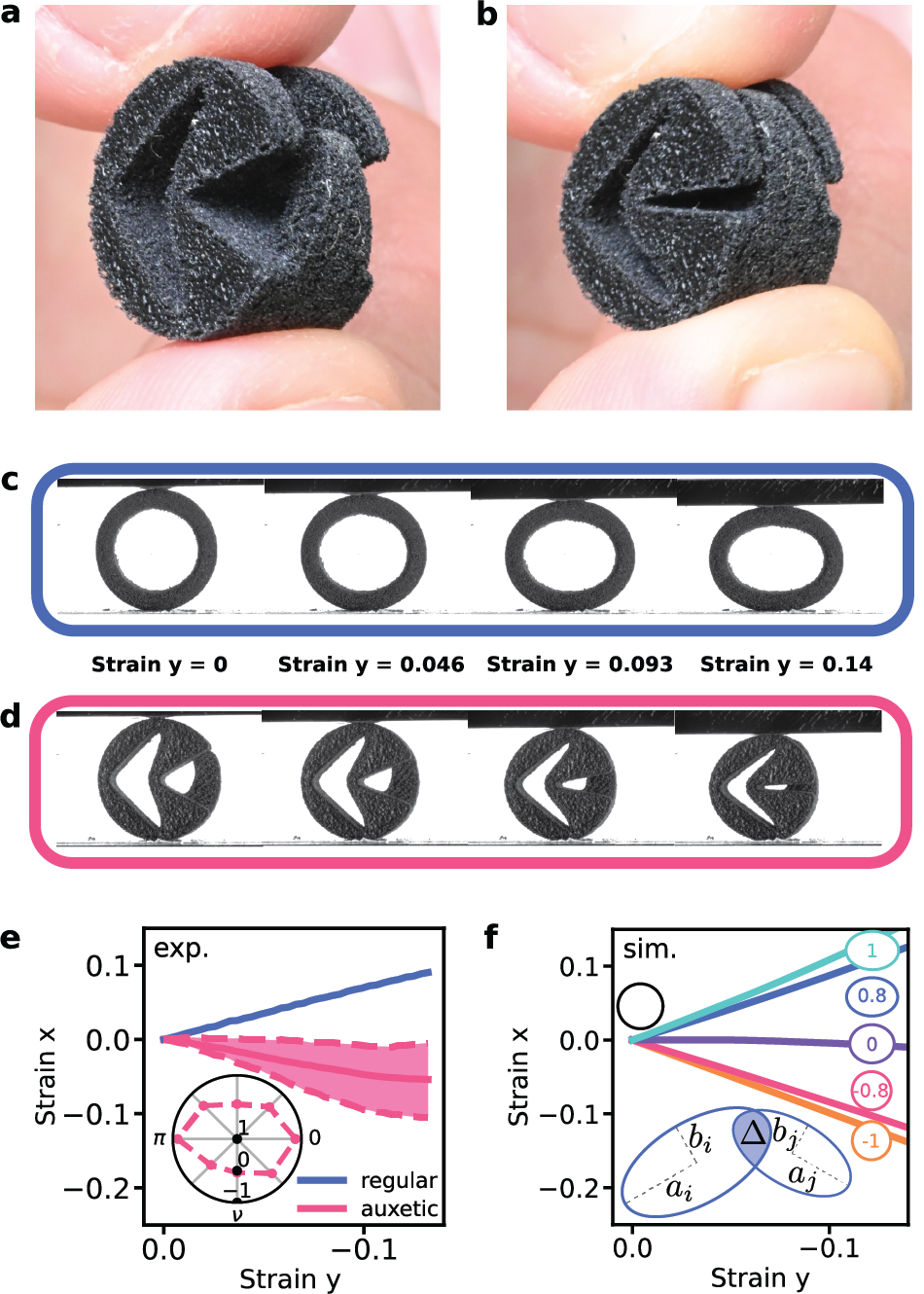}
\caption{\textbf{Auxetic metagrains} 
\textbf{(ab)} Picture of an auxetic disk (a) at rest, (b) under manual compression.
\textbf{(cd)} Four stages of compaction of a regular grain (c) and an auxetic grain (d).
\textbf{(e)} Plot of the strain in x-direction against the strain in y-direction of both a regular and an auxetic grain in experiment (flagged with 'exp.'); inset: Polar plot of the Poisson's ratio of the auxetic grain for eight angular orientations.
\textbf{(f)} Plot of the strain in x-direction against the strain in y-direction of simulated grains with Poisson's ratios ranging from -1 to +1 (flagged with 'sim.'); inset: sketch of the degrees of freedom in the simulated grains. See Appendix \ref{Num_model} and Supplementary Video 1 for more information.}
\label{fig:1}
\end{figure}

Mechanical metamaterials are carefully engineered architectures that exhibit tunable properties which surpass that of their constituents~\cite{BertoldiReview}.
Importantly, these metamaterials are solids with a well defined configuration of reference. 
In stark contrast, granular materials are not tunable and don't have a well defined configuration of reference. Their jamming and flowing properties are instead entirely controlled by the physics of contact and friction~\cite{jaeger1996granular, deGennesGranular,
van2009jamming, ZhangJamming, DuringJammingTransition, CoordinationNumberDuring}. The rapidly growing field of granular metamaterial precisely attempts to make the mechanical properties of granular material tunable~\cite{athanassiadis2014particle,miskinNatMat,JammingApplicationReview,DaraioStructuredFabrics,
Pashine2023, ZhangFollowUp,gaspar2010granular}.

Granular metamaterials with complex shapes have thus been introduced~\cite{miskinNatMat,athanassiadis2014particle, JammingApplicationReview, Tetrapods, Architecture, Hygroscopic,wang2020shear,Staples,DaraioStructuredFabrics, particleShapeModeling, DimersAndEllipses} that enhance interlocking and have been shown to enjoy additional mechanical stability.
Yet, in many applications, granular materials need to flow. So instead of amplifying friction and promoting jamming, an important goal is to create granular metamaterials that can flow more easily. To meet this challenge, we take inspiration from flexible metamaterials, which use flexible building blocks to exhibit shape-shifting and superior damping performances~\cite{BertoldiReview, metamaterial_applications, shockdampeningDavid, metamaterialshockdampeningreview}. The most basic flexible metamaterials are perhaps those that exhibit negative Poisson's ratio, by leveraging internal counter-rotations within their bulk~\cite{RonResch,BertoldiReview}. We therefore ask the simple question: what are the elastic and flow properties of a granular metamaterial whose constitutive grains have a negative Poissons' ratio?

Our design uses a re-entrant mechanism \cite{larsen1997design} that leads to reduction in the radius of the disk as the grain is compressed (Fig.~\ref{fig:1}ab, see Appendix \ref{grain_design} and Supplementary Video 1 for details). We benchmark this metagrain against a simple ring, which has be designed to exhibit the same stiffness and that instead expands laterally when compressed (Fig.~\ref{fig:1}c). Unlike the ring, the metagrain is not isotropic (Fig. \ref{fig:1}d), but the mechanism used by the metagrain consists of a single mode of deformation, which ensures that the grain is auxetic regardless of the direction of compression (Fig.~\ref{fig:1}e).

Next, we construct a minimal computational model to describe the auxetic disks and the rings. We remark that when compressed, the auxetic disks hardly change their shape and rather decrease their area (Fig. \ref{fig:1}d). In contrast, the rings adopt an elliptical shape while maintaining their volume approximately constant (Fig. \ref{fig:1}c). We therefore introduce a potential energy for each grain with only three main energy components; an inter-particle repelling energy, an area restitution energy and a shape restitution energy (see Appendix \ref{Num_model} for details):
\begin{align}
    V = \sum_{<i,j>}\frac{k}{2}\Delta_{ij}^{2} + \sum_{i}\frac{k_{A}}{2A_{i}^0}(A_i - A_{i}^0)^2 + \sum_i\frac{k_s}{2}\epsilon_i^2.
\end{align}
The first component is proportional to the overlap between particles $\Delta_{ij}$. The second component is the energy from changing each particle's area, $A_i = \pi a_i b_i$, with respect to the particle’s initial area $A_i^0$ where $a_i$ and $b_i$ denote the axes of the elliptical particle. Lastly, the third component is the energy required to change the particle's shape, $e_i= a_i - b_i$, considering an initial circular shape (Fig.~\ref{fig:1}f-inset). This distinctive model is reminiscent of vertex models that describe foams and tissues~\cite{VertexFoamsEpithelia, VertexNonaffine, VertexTransition} where area and shape play a crucial role, yet is designed to describe deformable grains instead of closely packed foams.

We performed quasistatic frictionless numerical simulations of the rings and auxetic disks. We work in the limit of small overlaps between particles $(k_A+k_s)/k \rightarrow 0$. In this limit and at small strains, the Poisson's ratio of the grains is given by  $\nu=\frac{k_A-k_s}{k_A+k_s}$. Thus by changing the area and shape stiffnesses correspondingly, we can tune the particle's Poisson's ratio from $+1$ to $-1$, which is indeed confirmed by simulations (Fig.~\ref{fig:1}f). Even though the model is quite simple in nature, it provides an accurate description of the experiments.

\begin{figure}[t!]
\centering
\includegraphics[width=1.0\linewidth]{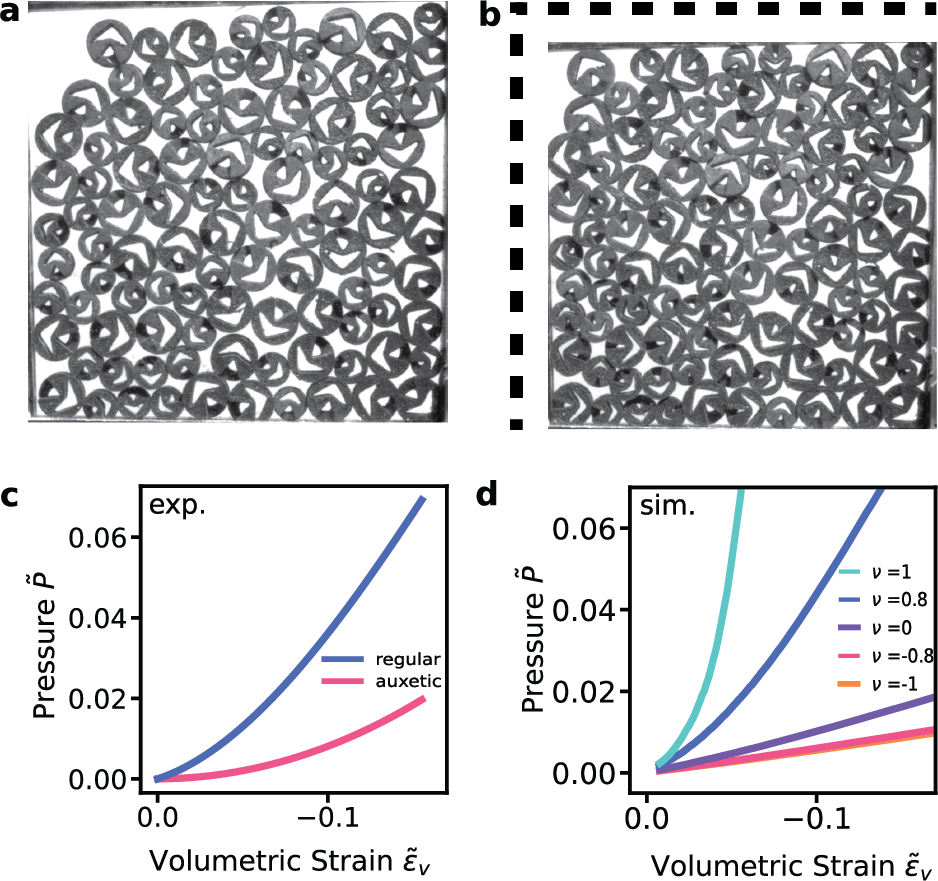}
\caption{\textbf{Auxetic packings are more compressible.} 
\textbf{(ab)} Equi-biaxial compression of a bi-disperse packing of 100 auxetic metagrains in the initial (a) and final (b) configuration. \textbf{(cd)} The pressure plotted against the strain of both the regular and an auxetic packings in the bi-axial compression in both experiment \textbf{(c)} and simulation \textbf{(d)}. The pressure is normalized for the grain stiffness (see Appendices \ref{guillotine_experimental}, \ref{guillotine_numerical}, \ref{Normalization_experimental} and Supplementary Video 2 for more information).
}
\label{fig:2}
\end{figure}

\begin{figure*}[t!]
\centering
\includegraphics[width=1.0\linewidth]{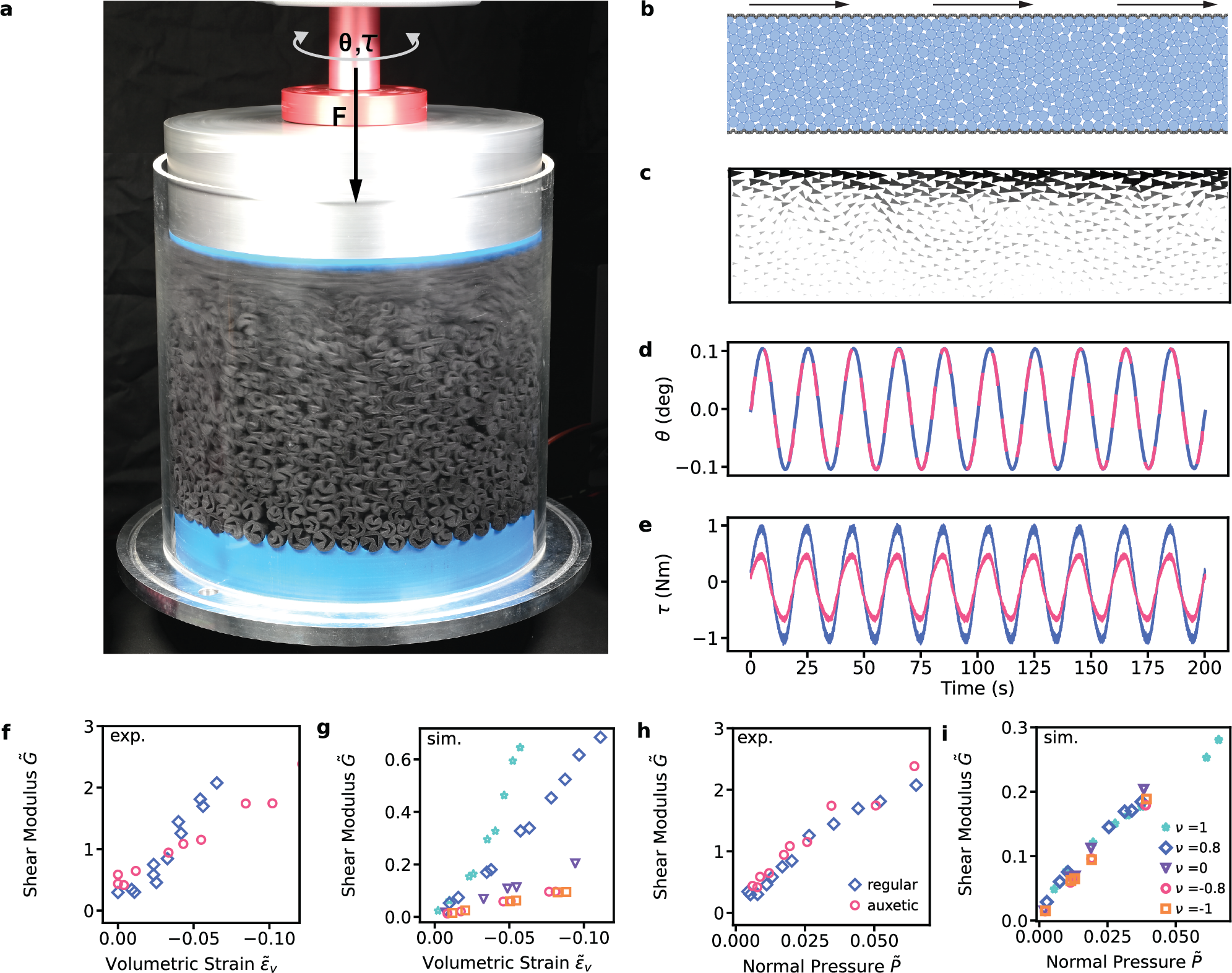}
\caption{\textbf{Auxetic packings have a smaller shear modulus.} \textbf{(a)} Photograph of a 2D Couette shearing experiment with an eight seconds shutter speed. A vertically imposed strain and rotational displacement ($\theta$) introduce a normal load ($F$) and torque ($\tau$) respectively.
\textbf{(bc)} Representation of a numerical shearing simulation of both granular location (b) and displacement field (c). The left and right boundaries are periodic similar to the experimental setup. The frictional boundary condition at the top is moving to the right.
\textbf{(de)} A small cyclic strain experiment with the imposed angle (c) and measured torque (d) against time for both the metagrains (pink) and the rings (blue) at a volumetric strain of 0.002.
\textbf{(fg)} Plot of the shear modulus in experiments (e) and simulations (f) vs. the imposed volumetric strain. 
\textbf{(hi)} 
Plot of the shear modulus in experiments (g) and simulations (h) vs. the measured averaged normal pressure (see Appendices \ref{shear_setup_experimental}, \ref{shear_setup_numerical}, \ref{Normalization_experimental} and \ref{Normalization_numerical} for details).
While experiments and simulations show trends in qualitative agreement, the shear modulus is much smaller in the simulations and vanish at vanishing pressure whereas they hit a constant value in the experiment. This is because we have not included friction and gravity in the model, which are known to enhance shear stresses~ \cite{van2009jamming}. See Supplementary Video 3 for more information.
}
\label{fig:3}
\end{figure*}

Now that we have a full description of a working metagrain, it is reasonable to wonder what would happen to a collection of them. Granular materials are well known to have a relatively larger bulk modulus and a relatively smaller shear modulus at the onset of rigidity~\cite{van2009jamming}. Here  we find that making the grains auxetic qualitatively changes this traditional picture.

We first measure the response of auxetic granular packings to equi-biaxial compression (Fig. \ref{fig:2}ab and Supplementary Video 2)---see Appendices \ref{guillotine_experimental} and \ref{guillotine_numerical} for details. It is clearly visible that the disks all shrink as the packing is being compressed. This facile compression of individual metagrains induces a much slower increase of pressure in comparison to that of the rings (Fig. \ref{fig:2}cd), which in contrast all tend to deform into ellipses that resist more compression (see Appendix \ref{guillotine_experimental} and Supplementary Video 2). This reduced bulk modulus was to be expected: each metagrain is itself more compressible, so is the packing.

We now move to a second important quantity that characterizes the elasticity of jammed packings: the shear modulus. To do this, we measure the shear rheology of packings under oscillatory shear in an custom-built Couette shear cell setup (Fig. \ref{fig:3}a, see Appendix \ref{shear_setup_experimental} and Supplementary Video 3 for more details) and by using simulations in a periodic box (Fig. \ref{fig:3}bc), at imposed confinement. 
In this regime, when subjected to small sinusoidal oscillations of magnitude $\theta=0.1\deg$, or equivalently, a shear strain of $0.002$  (Fig.~\ref{fig:3}d), the
packings respond with linear oscillations in the torque (Fig.~\ref{fig:3}e), hence the rheological response is linear (see Appendix \ref{Normalization_experimental} for details). Strikingly, we find that at large confinements, the shear torque of the metagrains is twice as low as that of the rings. This trend is confirmed by measurements of the shear modulus over a large range of confinements (Fig.~\ref{fig:3}fg): the shear modulus increases with confinement faster for the rings than for the metagrains. This is surprising from the perspective of 2d isotropic continuum elasticity, where the bulk and shear modulus are given by $K=E/(2-2\nu)$ and $G=E/(2+2\nu)$, where $E$ is the Young's modulus. Hence, since both types of grains have the same stiffness (see Appendix \ref{grain_design} for more details) and that the metagrains have a lower bulk modulus, one would have expected the packing of metagrains to exhibit a larger shear modulus. 

To elucidate this issue, we plot the shear modulus against the confining pressure instead of the volumetric strain (Fig.~\ref{fig:3}hi). Surprisingly, we find that the data points collapse onto a master curve that grows sublinearly with the confining pressure. This surprising collapse is due to two competing effects in the packing geometry and Poisson effect. Packings of auxetic grains make up a denser contact network at a given pressure (see Appendices \ref{Poissons_effect} and \ref{poisson_effect_numerical}), which is expected to increase the elastic modulus. Yet, their negative Poisson's ratio tends to decrease stresses in the direction that is transverse to that of the compressive stresses, hence compensating for the larger elasticity of the denser contact network. The cause is the purely repulsive nature of the interaction forces between particles, which is fundamentally different from ordinary elastic solids that admit both repulsive and attractive forces within their bulk. 

\begin{figure}[t!]
\centering
\includegraphics[width=1.0\linewidth]{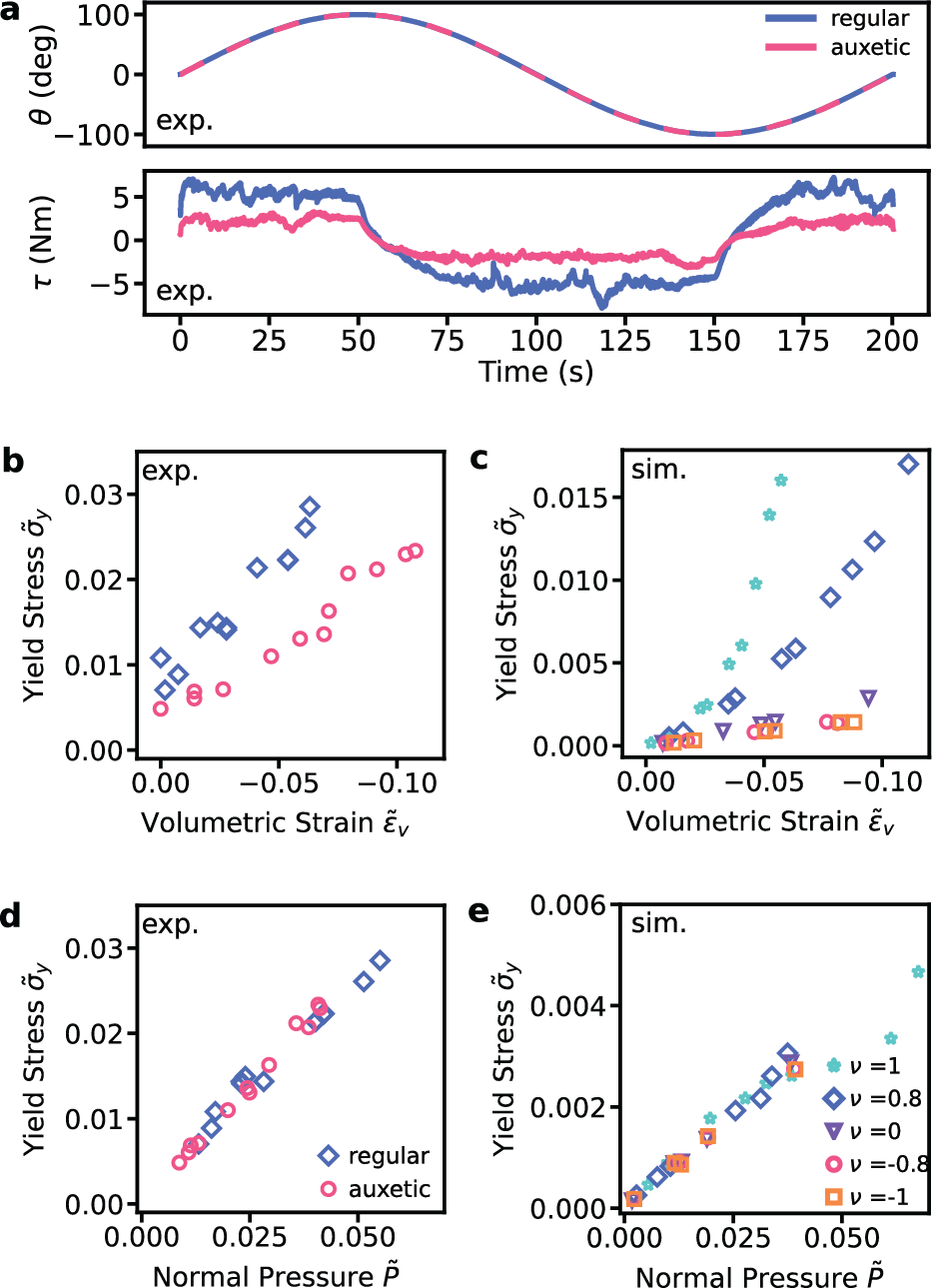}
\caption{\textbf{Auxetic packings flow more easily.}
\textbf{(ab)} A large cyclic strain experiment with the imposed angle equivalent to a shear strain of $2.00$ (a) and measured torque (b) against time for both the metagrains (pink) and the rings (blue).
\textbf{(cd)} Plot of the yield stress in experiments (c) and simulations (d) vs. the imposed volumetric strain. 
\textbf{(ef)} 
Plot of the yield stress in experiments (g) and simulations (h) against the measured averaged normal pressure. Again, the experiments and simulations show trends in qualitative agreement, but the yield stress is much smaller in the simulations and vanish at vanishing pressure whereas they hit a constant value in the experiment. This is again due to the simplicity of the model. In the flowing regime, this leads to the appearance of a shear band as can be seen in Fig.~\ref{fig:3}a, which is not present in the simulations. See Appendices \ref{shear_setup_experimental}, \ref{shear_setup_numerical}, \ref{Normalization_experimental}, \ref{Normalization_numerical} and Supplementary Video 4 for details}
\label{fig:4}
\end{figure}

\begin{figure}[t!]
\centering
\includegraphics[width=1.0\linewidth]{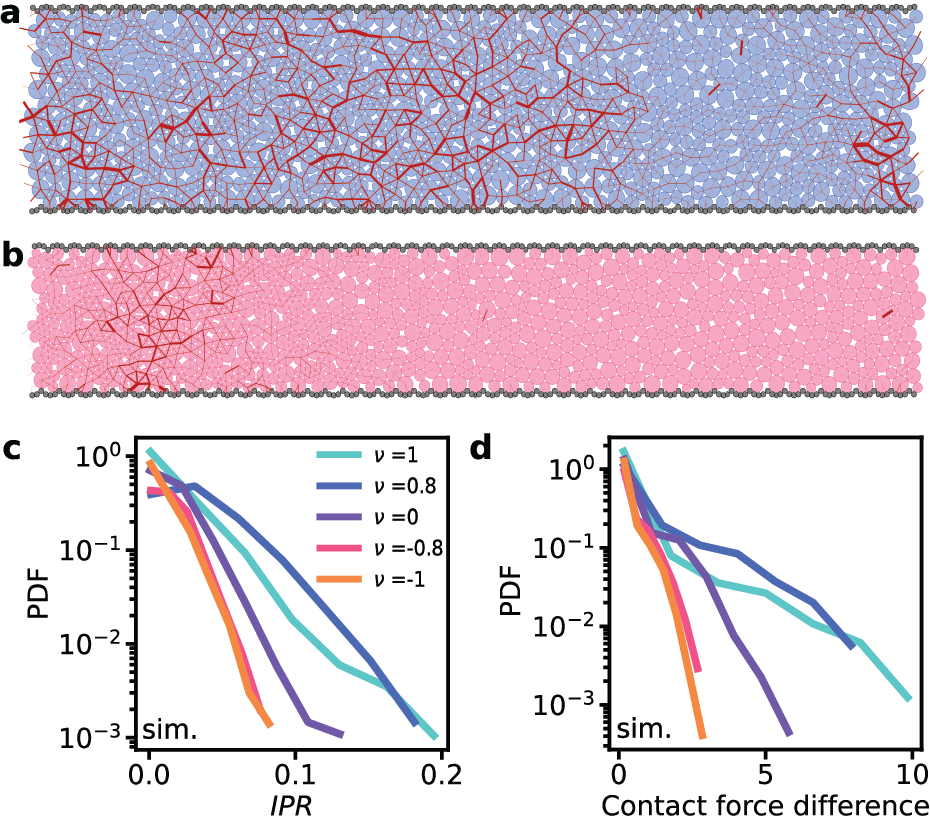}
\caption{\textbf{Auxetic packings  fluctuate less.} \textbf{(ab)} The difference in inter-particle forces between two frames during an typical avalanche in a numerical simple shear simulation with both a regular (a) and auxetic packing (b). Both are compressed in volume in order to obtain the same normal pressure (see Supplementary Video 5 for more information). \textbf{(cd)} The probability density function (PDF) of the inversed partition ratio (c) and of the inter-particle contact forces as a consequence of an avalanche in numerical simulations. An inversed partition ratio of $1$ indicate that all particles  are involved in an avalanche---$1/N$ if it is only single particle.
}
\label{fig:5}
\end{figure}

\begin{figure*}[t!]
    \centering
\includegraphics[width=0.8\linewidth]{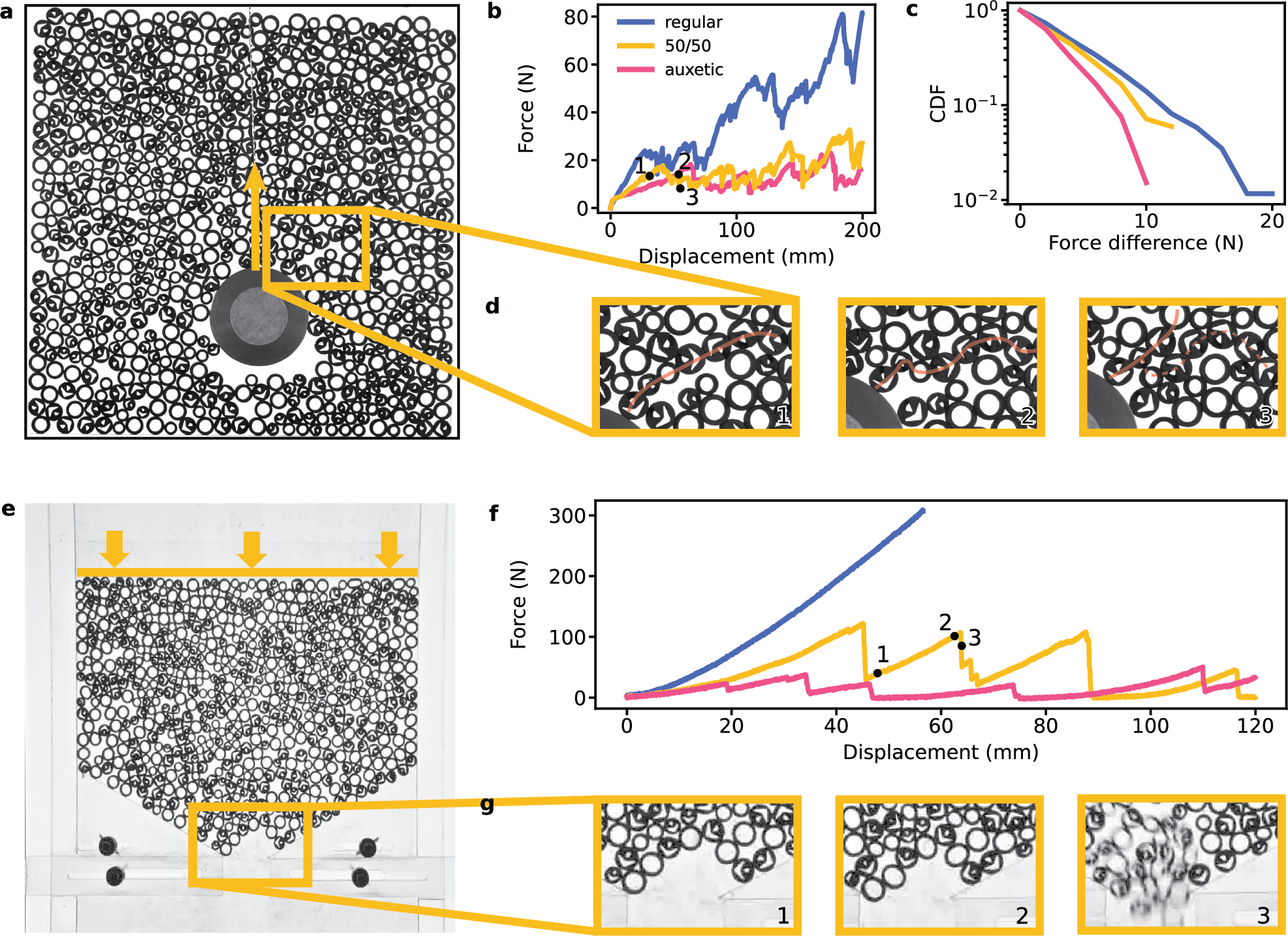}
\caption{\textbf{Typical enhanced granular flows in confinement} \textbf{(a)} Experimental setup of an 8cm intruder in a confined equally mixed packing of auxetic and regular grains. The intruder is pulled through the medium with a metal wire. \textbf{(b)} Force curves of the displacement of the intruder in regular, auxetic and a mixed packing. \textbf{(c)} The cummulative distribution function (CDF) of events where a force chain breaks with the resulting difference in force. \textbf{(d)} Three consecutive zoom ins on a typical breaking of a force chain, indicated in red. The three stages represent the force chain with low force, (1) during a build-up of force (2) and after the release of force by breaking the force chain (3). The auxetic grains in the chain have shrunk in panel (2) and expanded after the breaking of the force chain \textbf{(e)} Experimental setup of the compression of an equal mixture of auxetic and regular grains in an hourglass configuration (semi-confined). \textbf{(f)} Force-displacement curve of compressed packings of regular, auxetic and an equal mixture of grains in an hourglass configuration. \textbf{(g)} Zoom-ins of three different stages of the breaking of a force chain in an equal mixture of auxetic and regular grains. The stages represent the arc under little compression (1), high compression (2) and right after the break of the chain (3). See Supplementary Video 6 for more information.}
\label{fig:6}
\end{figure*}

This very absence of attractive forces is also the reason why granular materials yield and flow more easily. 
We demonstrate in the following that the same competition between contact network and Poisson effect not only affect the elastic response of granular metamaterials but also their rheology.
We perform oscillatory rheology on our packings, 
but this time at large strains $\theta=100~\deg$, or equivalently a shear strain of $2.00$ (Fig.~\ref{fig:4}a top). The shear stress this time plateaus at well defined yield stress values, only to change sharply when the direction of shear is reversed (Fig.~\ref{fig:4}a bottom). Crucially, in collections of metagrains, the yield stress is twice as small, fluctuates less (Fig.~\ref{fig:4}a bottom), and increases much less quickly with confinement (Fig.~\ref{fig:4}bc). To better understand where this strong difference comes from, we again plot the yield stress against pressure and find as above that the data points collapse onto a master curve slower with confinement (Fig.~\ref{fig:4}de). Strikingly, the granular media have the same macroscopic friction coefficient $\tau/P$, whether they are made of auxetic particles or not. As is the case for its elasticity, the friction coefficient of granular media is also understood to be primarily controlled by the network of contact forces ~\cite{lerner2012unified}, which is denser for packings of auxetic grains (see Appendices \ref{Poissons_effect} and \ref{poisson_effect_numerical}) and is hence expected to be harder to shear. Yet again, auxetic particles shrink under compression, which reduces the overall stresses hence compenstates for the denser contact network.

Although the average rheology is the same at imposed pressure, the fluctuations dramatically differ. Our simulations show that packings of auxetic particles have much more reduced and localized force fluctuations than that of ordinary packings~(Fig.~\ref{fig:5}ab). 
In particular, we see that when an avalanche---viz. a sudden drop in shear stress (see Appendix \ref{shear_setup_experimental})---occurs, less particles are involved, as measured by the inversed partition ratio (Fig.~\ref{fig:5}c). Also, the inter-particle forces of the grains that are involved typically decreases less than in a regular packing (Fig.~\ref{fig:5}d). Auxetic grains thus seem to strongly facilitate the failure of force chains (see Supplementary Video 5 for more information).

In summary, the Poisson's ratio has no sensible effect on the average elastic and flow response to shear at imposed pressure, but it has at imposed volume. In such case, it is the enhanced compressibility of the metagrains that makes granular metamaterials considerably easier to shear. The auxeticity of the particles is also what considerably reduces the fluctuations when the granular metamaterials flow. 

Granular metamaterials will hence have the most dramatic effect in confined geometries that involve strong fluctuations. We conclude this paper by illustrating two such situations, that of a flow through a constriction and of an intruder moving through a granular material.

In the first situation, the box is closed and the intruder is $5.7$ times larger than the largest grains (Fig.~\ref{fig:6}a), hence the flow is confined and the small separation of scales between the intruder and the grains will induce large fluctuations. We measure that it is easier to traverse through a packing of metagrains or a mixture of metagrains and rings compared to a packing of rings only (Fig.~\ref{fig:6}b). The size of the avalanches is also much reduced (Fig.~\ref{fig:6}c). Although it shows larger fluctuations, it is remarkable that the mixture shows almost the same drag force than the pure packing of metagrains. 
This implies that adding of a fraction of metagrains to a regular packing is sufficient to break force chains. This is confirmed by visual inspection of such breaking event, where indeed we see that it is the compression of a metagrain that breaks up the force chain (Fig.~\ref{fig:6}d) -- See Supplementary Video 6 for more information.

In the second situation---perhaps the oldest granular problem~\cite{FirstGranularFlow,franklin1955flow, BEVERLOO1961260}---the role of fluctuations is paramount as the flow of particles is unexpectedly interrupted due to the sudden formation of a stable arc of contacting grains in front of its orifice. In our demonstration the size of the orifice is $3$ times that of the largest grains and we push from the top by a piston, hence such flow shows strong fluctuations (Fig.~\ref{fig:6}e). While the rings clog, which leads to a continued rise in the force exerted by the piston, both pure packing of metagrains and mixture flow (Fig.~\ref{fig:6}f), with again the mixture requiring larger avalanches to flow. We attribute this large build up of force to the lesser compressibility of rings and the smaller change of failure of an arches comprising less auxetic particles. This can be readily seen in Fig.~\ref{fig:6}g, where only two auxetic particles take part in the arch that momentarily clogs the hopper -- See Supplementary Video 6 for more information.

To conclude, we have added auxeticity to the toolbox of granular metamaterials and we have shown that the granular metamaterials can be made to be more compressible, softer to shear and to yield more easily with less fluctuations.
Exciting open questions ahead are how to extend the concept beyond 2d auxetics, for instance in three dimensions and for more generic mechanical responses beyond auxeticity such as complex shape-changes and bistability. We envision the enhanced properties of granular metamaterials to open avenues for granular transport~\cite{KeyProblemsFlow,MuOfI,SolutionGanularFlow, GranularFlowVertical},  energy absorption~\cite{wang2021structured, ImpactForcePropagation, ImpactProjectile, ForceLawImpact, TruckEscapeRamps} and soft robotics~\cite{brown2010universal, RoboticArm}.

\emph{Acknowledgments.} 
 We thank Daniel Bonn, Joshua Dijksman and Heinrich Jaeger for insightful discussions and suggestions, Antoine Dop and Mees Wortelboer for preliminary experimental and numerical explorations and Clint Ederveen Janssen, Daan Giesen, Gerrit Hardeman and Sven Koot for technical assistance. We acknowledge funding from the European Research Council under grant agreement 852587 and from the Netherlands Organisation for Scientific Research under grant agreement VI.Vidi.213.131.3 

All the codes and data supporting this study are available on the public repository  \url{https://doi.org/10.5281/zenodo.8405504}.
\bibliographystyle{naturemag}

\bibliography{Reference}

\clearpage
\setcounter{equation}{0}
\renewcommand{\theequation}{A\arabic{equation}}%
\setcounter{figure}{0}
\renewcommand{\thefigure}{A\arabic{figure}}%
\renewcommand{\figurename}{\EDF}%

\onecolumngrid
\begin{appendix}

\setcounter{tocdepth}{2}

\tableofcontents
\newpage
\section{Designing auxetic and regular grains} \label{grain_design}

\begin{figure*}[h]
    \centering
\includegraphics[width=1.0\linewidth]{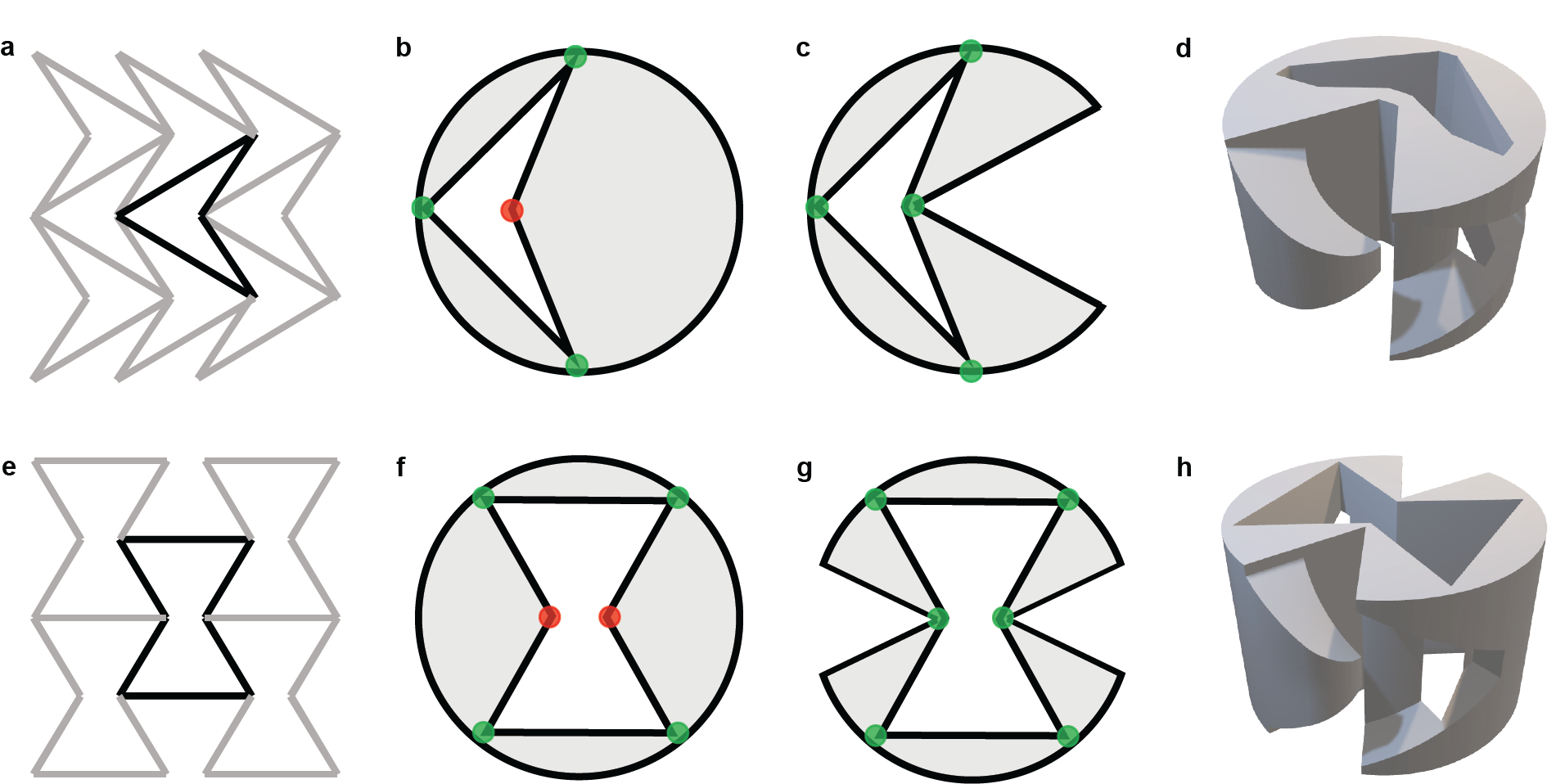}
\caption{\textbf{Auxetic grain design scheme. (a)} Re-entrant auxetic structure. \textbf{(b)} Grain with a unit cell hole. Flexible hinges are highlighted with green dots, where as rigid hinges are highlighted with red dots. \textbf{(c)} 2D Auxetic grain design. There are no more rigid hinges due to the cutout from the perimeter. \textbf{(d)} 3D auxetic grain design with obstruction to prevent interlocking between grains. \textbf{(e)} We consider a second re-entrant auxetic structure. \textbf{(f)} Analogue to our first design, we obtain a grain with a unit cell hole. \textbf{(g)} The rigid hinges are made flexible with cutouts. \textbf{(h)} With obstructions to prevent interlocking, we find a second auxetic grain design.}
\label{SI:scheme}
\end{figure*}

In this section, we expand on the design of the grains. The grains used throughout the paper are inspired by a known auxetic re-entrant structure (fig \ref{SI:scheme}a) \cite{larsen1997design}. The structure is a repetitive pattern of the highlighted unit cell. The shape of the unit cell forms a hole in our granular design (fig \ref{SI:scheme}b). For the hole to deform, we require that the corners are very flexible and that the interconnecting bars are rigid. When the corners are close to the perimeter of the grain, the hinges are flexible. We see that in this design, three out of four hinges are flexible, highlighted in green. The rigid hinge is highlighted in red. We remove a triangular section of the grain to also allow the last hinge to become flexible (fig \ref{SI:scheme}c), thus making the grain auxetic. In contact with similar grains, however, there will be events of interlocking between grains. To prevent this from happening we added a tooth on the circular perimeter of our grain (fig \ref{SI:scheme}d), which acts as a barrier. An empty pocket in the grain sheathes the tooth and prevents self-contact when the grain is compressed. 

This is by no means the only auxetic grain design we can think of. We can create more auxetic grains with the same design scheme. A second example is shown in figures \ref{SI:scheme}e - \ref{SI:scheme}h. The desired granular properties can be altered by choosing the best corresponding re-entrant structure. We choose to work with the design depicted in figure \ref{SI:scheme}d as it displays, apart from a negative Poisson's ratio, only one degree of freedom. This implies that a change in one of the hinges determines how the other hinges will deform. This ensures that the grain is robust under both compression and shear and, quite importantly, behaves predictable in an ensemble where each grain is in contact with more than two other grains.\\

\begin{figure}[h]
    \centering
\includegraphics[width=0.5\linewidth]{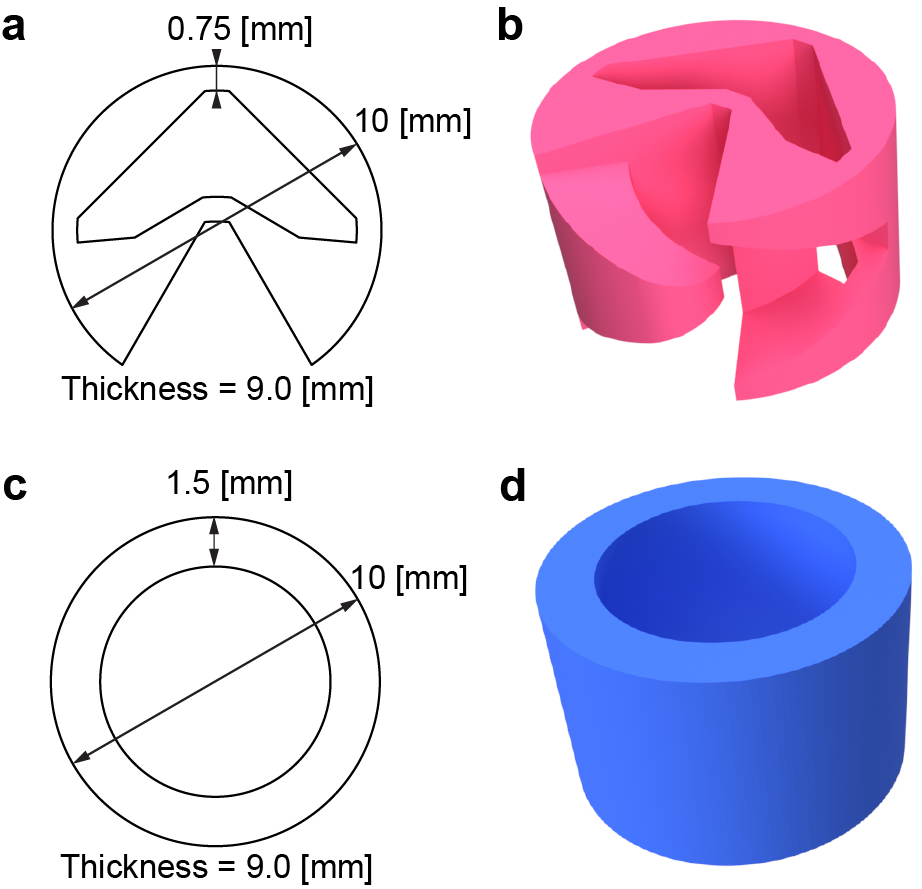}
\caption{\textbf{Granular design parameters} \textbf{(a)} A 2-dimensional top view of the auxetic grain design. \textbf{(b)} A 3-dimensional view on the auxetic grain model. The added obstruction and cutout prevent interlocking between grains. \textbf{(c)} A 2-dimensional top view of the regular grain design. \textbf{(d)} A 3-dimensional view on the regular grain model}
\label{SI:design}
\end{figure}

For the fabrication of the grains we used a Selective Laser Sintering printer (Sinterit Lisa) that prints a thermoplastic polyurethane with a Young's modulus of 3.7 MPa (FlexaBlack). The powder is sintered layer by layer to form an elastic-like solid design. The selected layer height of the printer is 200 microns. To avoid any issues with the printability, we modified the hinge sizes from our conceptual design to become a slender beam of $0.75$mm thickness (fig \ref{SI:design}a). A 3-dimensional model of the auxetic grains is shown in figure \ref{SI:design}b. The regular grain design is an extruded ring. The thickness of $1.5$mm is designed such that the stiffness of the auxetic and regular grains are comparable (fig \ref{SI:design}c). A 3-dimensional model of the regular grain is shown in figure \ref{SI:design}d. Both the auxetic and the regular grains have a height of $9$mm. The grains are printed in the magnification as depicted in figure \ref{SI:design} as well as in a magnification of 1.4 times this representation. The height is unchanged in this magnification. Our experiments are performed with packings of an equal mixture of both sizes of grains. This avoids meaureable artefacts as a result of crystallization within the packing.

\begin{figure}[h!]
    \centering
\includegraphics[width=0.5\linewidth]{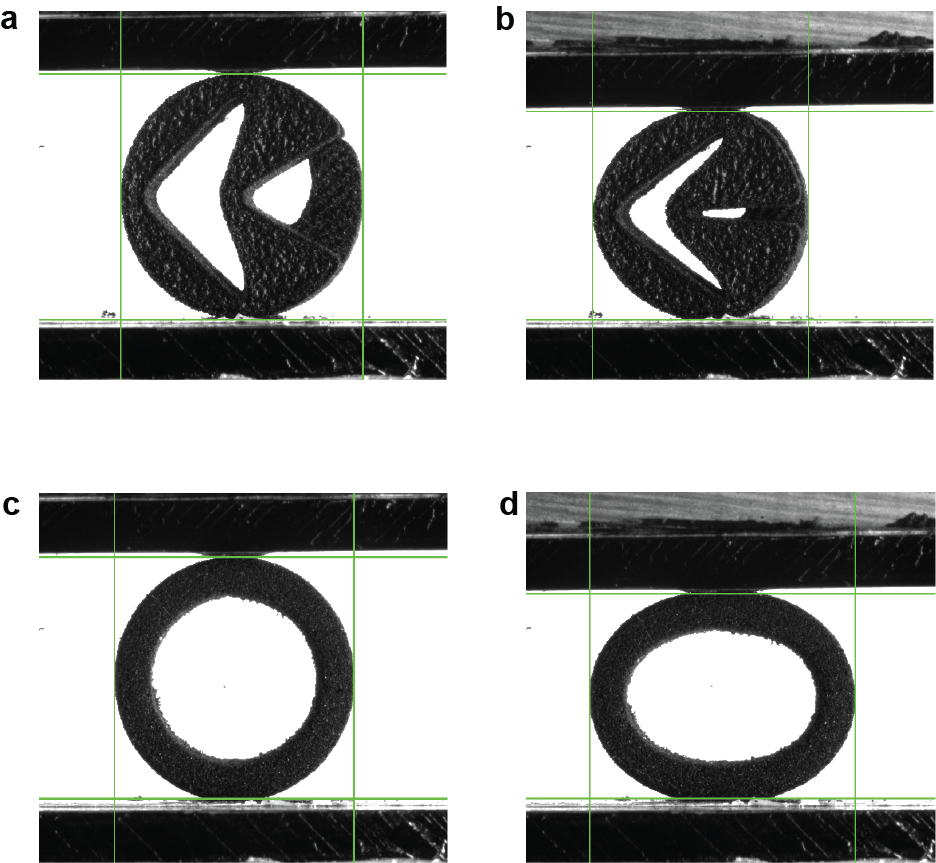}
\caption{\textbf{Tracking of the grain's Poisson's ratio.} \textbf{(a)} An undeformed auxetic grain in the orientation we benchmark in the main text as '0 degrees'. The position of the green lines is detected and consequently used for determining the strain in both x and y directions. \textbf{(b)} An auxetic grain deformed by a planar compaction. \textbf{(c)} An undeformed regular grain. \textbf{(d)} A regular grain deformed by a planar compaction.}
\label{SI:poisson}
\end{figure}

The Poisson's ratio of the individual grains is measured from a planar compression in a uni-axial single column testing device (Instron 5943 series) with a 500N load cell. The frames are recorded with 14 fps and a resolution of 2700 px $\times$ 2400 px with a Basler Ace acA3800-14um camera. Since the auxetic grain has an anisotropic shape, we measure the Poisson's ratio under eight different orientations. Figure \ref{SI:poisson}a shows an undeformed auxetic grain in the orientation we refer to as '0 degrees'. With equally spaced intervals of 45 degrees counterclockwise we consecutively perform the planar compression eight times before we reach the initial rotation again. The compression is performed for three distinct grains. The grain size in y direction is imposed to be linearly decreasing. By finding the two best horizontal lines with the OpenCV library, we define a region of interest. This region of interest is binarized such that there is a clear distinction between the grain and the background. This region of interest now excludes the top and bottom part of the compression setup. The grain size in x direction is now measured by finding the first and the last column of the image where there are black pixels. 

 After compression, it is visible that the auxetic has shrunk in both directions (fig \ref{SI:poisson}b). The regular grain however expands in the x axis with a decreasing y axis (figs \ref{SI:poisson}c and \ref{SI:poisson}d). Since the regular grain is symmetric under rotations, the Poisson's ratio as given in main text figure \ref{fig:2}b is the average of three distinct grains. 

\begin{figure}[h]
    \centering
\includegraphics[width=0.5\linewidth]{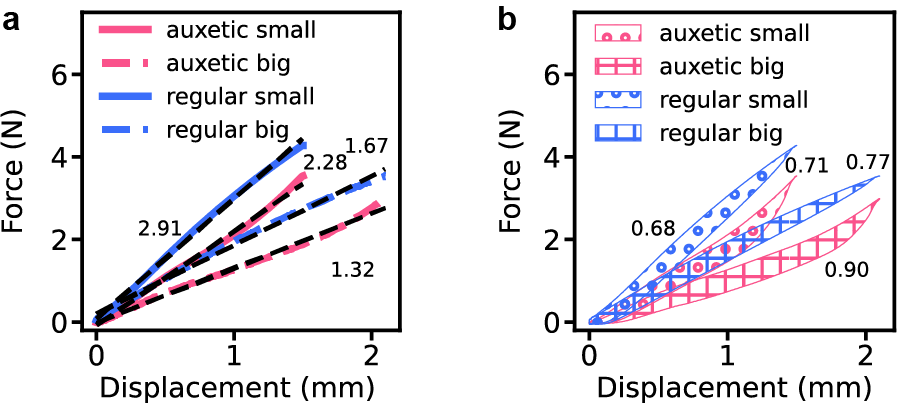}
\caption{\textbf{Single grain stiffness and dissipation.} \textbf{(a)} The force displacement curves for the big and small grains used in the bidisperse packings for both auxetic and regular grains. The lines are averages over three distinct grains and for the auxetic grains averaged over the eight measured orientations. The black dashed lines are linear fits to determine the stiffness of the grains which is used for the normalization of the experiments. \textbf{(b)} The full force displacements curves of the cyclic compression of auxetic and regular grains. The force difference for compression and decompression are highlighted with the shaded areas for each curve. The amount of dissipated energy (in Nmm) is shown written with the corresponding dissipation curve.}
\label{SI:grain_properties}
\end{figure}

The stiffness of the auxetic and regular grains are obtained from the same planar compression. After averaging the force displacement curves of the measured grains we find force displacement curves as shown in figure \ref{SI:grain_properties}a, where the measurements on the big grains are shown as solid lines and on the small grains as dashed lines. The dashed black lines are linear fits through the data. The values that next to the dashed black lines are the slopes of the linear fits. For the normalization of the experiments we have used the averaged stiffness of the big and small grains, $2290$N/m for the regular grains and $1800$N/m for the auxetic grains. By plotting not only the compression, but also the decompression of the same experiment, we obtain the force displacement curves shown in figure \ref{SI:grain_properties}b. The shaded areas correspond to the dissipated energies during a full cycle. The numbers next to the shaded areas correspond to the size of the shaded areas in Nmm. We observe that the dissipation between auxetic and regular grains is comparable, which indicates that this is not an effect that can lead to qualitative differences in packings of auxetic and regular grains.



\section{Experimental uniaxial setup} \label{guillotine_experimental}
\begin{figure}[h!]
    \centering
\includegraphics[width=0.5\linewidth]{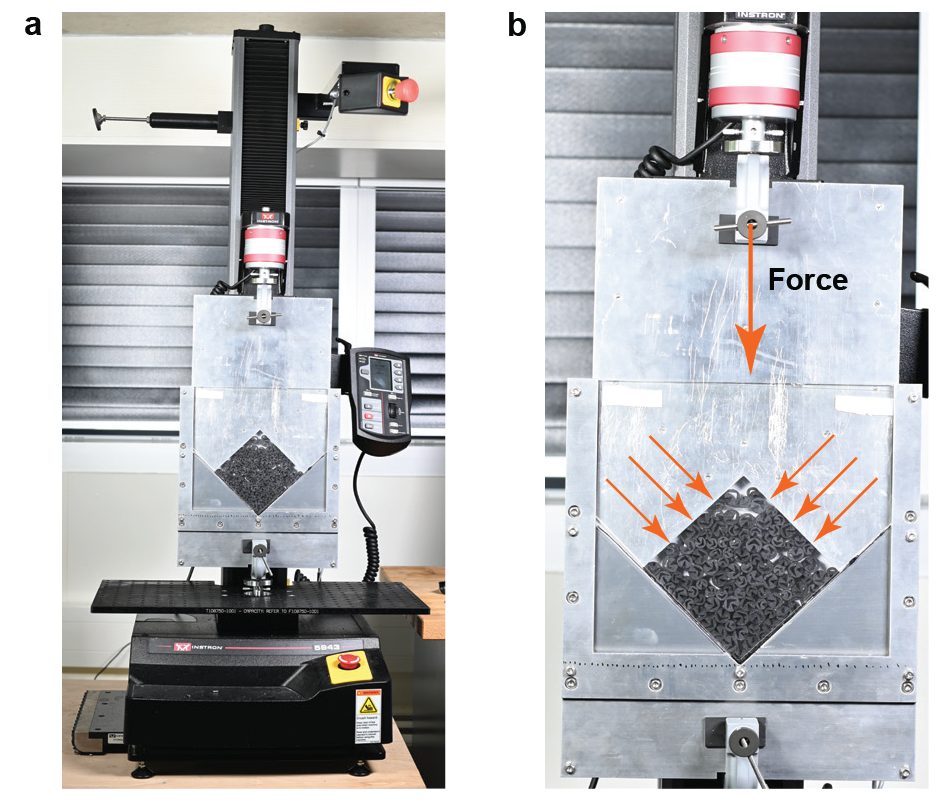}
\caption{\textbf{Bi-axial compression experiments.} \textbf{(a)} The bi-axial experimental setup mounted in a uni-axial tensile testing machine (Instron 6800 series). \textbf{(b)} A zoom in on the bi-axial experimental setup. A metal plate with a reversed v-shape is moved into a v-shaped construction containing grains. The force is therefore applied equally from two directions as indicated by the arrows.}
\label{SI:guillotine}
\end{figure}

The compressive behaviour of packings of grains has been tested in a uni-axial single column testing system (Instron 5943 series) with a 500N load cell. We used a custom-made aluminium V-shaped press used and described by Corentin Coulais et al. in their earlier work \cite{Coulais2018biaxial}. The shape ensures that the pressure on the packing is applied equally from two directions. A bidisperse mixture of 100 grains is randomly inserted after which it is compressed with a speed of 1 millimeter per second for a total distance of 150 millimeter. The full experimental setup is shown in figure \ref{SI:guillotine}a and a zoom in on the V-shaped press filled with grains in figure \ref{SI:guillotine}b.

\section{Experimental shearing setup} \label{shear_setup_experimental}
\begin{figure}[h]
    \centering
\includegraphics[width=1.0\linewidth]{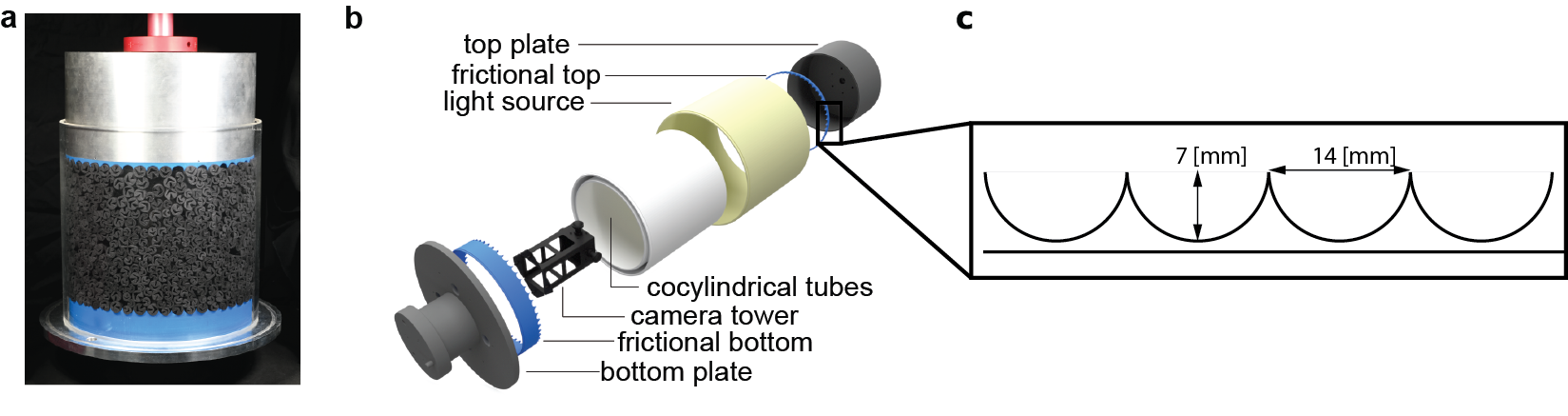}
\caption{\textbf{Shearing experimental setup.} \textbf{(a)} The home-built Couette setup in which we perform cyclic shearing experiments. \textbf{(b)} A construction view of a three-dimensional model of the shearing setup. A printed frictional wall is inserted both in a top and bottom plate. Two transparent cocylindrical tubes contain the grains. From the inside of the tubes, the grains are filmed with four cameras with fish-eye view. A cocylindrical light source is placed around the cylindrical grains setup. \textbf{(c)} Sketch of the profile of the frictional wall.}
\label{SI:Couette}
\end{figure}

The shearing behaviour of packings of grains has been investigated in a linear-torsion dynamic test instrument (Instron, Electropuls E3000 TT). We mount two concentrical plexiglass cylinders with diameters 220 and 250 millimeters respectively and thicknesses of both 5 millimeters to a metal disk which is attached to the base of our test instrument. A metal cylinder with a diameter of 235 millimeters is attached to the linear-torsion load cell. The metal cylinder perfectly fits in between the two plexiglass cylinders without friction. A 3D printed frictional wall is located both at the bottom plate in between the two cylinders and directly attached to the top cylinder. The wall has periodical asperities with a spacing of 14 millimeters, the same size as the big grains in equilibrium position. A photo of the full setup with grains is shown in figure \ref{SI:Couette}a. A 3-dimensional model with a construction view of the separated parts of the setup is shown in figure \ref{SI:Couette}b. A sketch of the frictional wall is depicted in figure \ref{SI:Couette}c.

\begin{figure}[h]
    \centering
\includegraphics[width=1.0\linewidth]{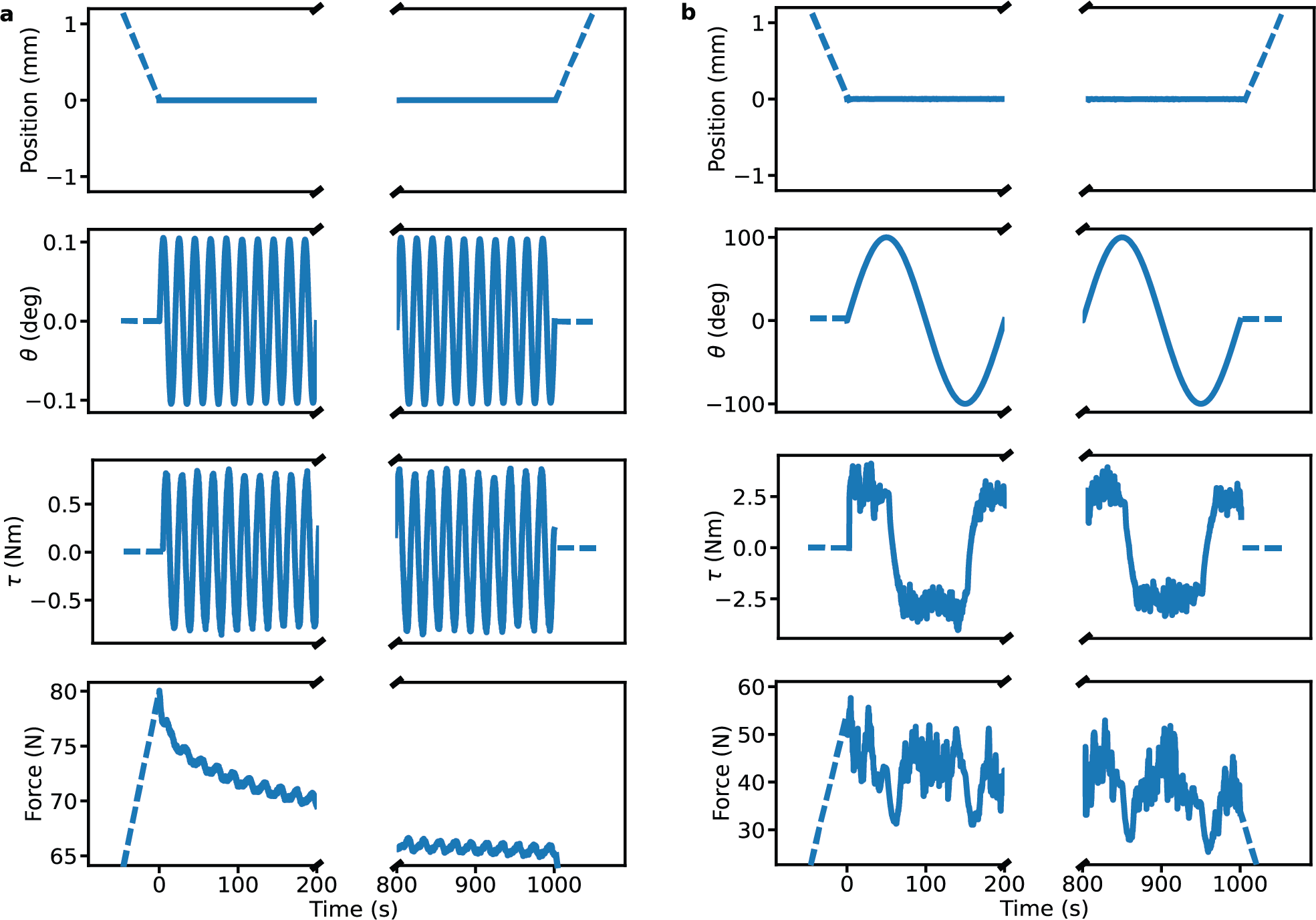}
\caption{\textbf{Shearing protocol.} \textbf{(a)} The cyclic shearing protocol of a typical small strain shearing experiment. The position of the frictional wall is lowered to an imposed value before the cyclic shearing is started. The position  is kept constant as the angle alters with an amplitude, $\theta$, of $0.1$ degree and a frequency of $0.05$ Hz. As a linear elastic response, we measure that the torque ($\tau$) fluctuates along with the angle. The force during the experiment decreases due to aging of the packing. \textbf{(b)} The cyclic shearing protocol of a typical large strain shearing experiment. The position of the frictional wall is lowered to an imposed value before the cyclic shearing is started. The position is kept constant as the angle, $\theta$, alters with an amplitude of $100$ degrees and a frequency of $0.005$ Hz. The torque, $\tau$, reaches a dynamic yield stress of which the signs changes upon a change in the shearing directin. The force during the experiment decreases slightly due to aging of the packing. }
\label{SI:Shearing_protocol}
\end{figure}

\begin{figure}[h]
    \centering
\includegraphics[width=0.5\linewidth]{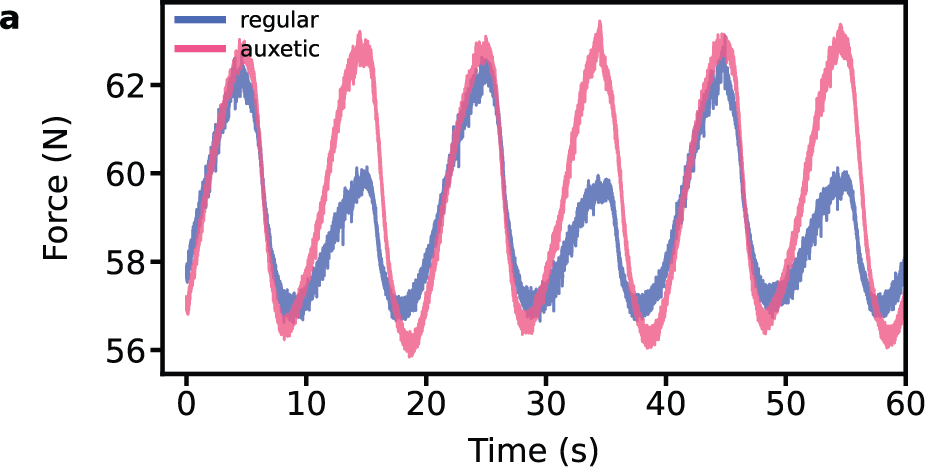}
\caption{\textbf{Typical dilation measurement} \textbf{(a)} The force as a function of time during a cyclic shearing experiment with an amplitude of $2.0$ degrees. This amplitude is larger than the $0.1$ degrees from which we extracted the linear elastic response, but low enough to avoid particle reorganisation.}
\label{SI:Dilation}
\end{figure}


After randomly inserting a bidisperse packing of $800$ grains, we conduct shearing experiments in both the reversible and irreversible regime. For the small strain reversible experiments, we first confine the packing by lowering the metal cylinder. Afterwards, we perform a cyclic shearing with an amplitude of $0.1$ degrees and a frequency of $0.05$ Hz for $50$ cycles. The protocol of these experiments is shown in figure \ref{SI:Shearing_protocol}a, where the position. The $0.1$ degree amplitude is approaching the limit of our device to perform accurate measurements. The frequency is selected to obtain experimental data at a reasonably quick pace whilst not moving so quick that there is a significant influence of the shear rate. The packing ages significantly during the experiment, which is clearly visible from the force data . The analysis on the shear modulus has been performed only on the last $10$ cycles of the experiment to avoid the effect of ageing as much as possible. In these last $10$ cycles we observe that the force has only a marginal time dependence

For the large strain irreversible experiments, we also first confine the packing. Here, we perform a cyclic shearing with an amplitude of 100 degrees and a frequency of 0.005 Hz for 5 cycles. Figure \ref{SI:Shearing_protocol}b shows the protocol for the large angular shearing experiments. The large angular rotation ensures that the grains are irreversibly moving. The amplitude is lowered with respect to the small angular shearing experiments to avoid unintended effects due to high shear rates. The aging in these irreversible experiments is typically less persistent due to the many reorganisations in the experiment. We therefore measure the yield stress in the last of $5$ cycles. Between two shearing experiments the grains are manually stirred to avoid crystallization effects that lead to denser packings.

Apart from the small angular shear strain and large angular shear strain experiments reported in the main text, we also performed intermediate angular shear strain experiments with an angle of $2.0$ degrees to probe the non-linear elastic response of auxetic and regular granular packings. This angle was chosen since irreversible particle just does not occur yet whilst the non-linear response would be maximally expressed. Our interest was predominantly guided towards the Reynolds dilatancy. For the volume controlled experiments this effect is observed as the increase of normal force when the packing is sheared out of its equilibrium. A typical force curve for cyclic shearing over time is shown in figure \ref{SI:Dilation}a for both a packing of auxetic and regular grains. From our experiments, we could not distinguish a strong difference in this non-linear effect for the two different Poisson's ratio grains. The difference in peaks for the regular packing is a result of pre-stress in the packing at the point defined as 0 degrees of shear. Further experiments with more control for the initial packing conditions will be necessary to obtain a more conclusive understanding of the non-linear effects of auxetic grains in a granular packing.

\section{Poisson's effect} \label{Poissons_effect}

\begin{figure}[h]
    \centering
\includegraphics[width=0.5\linewidth]{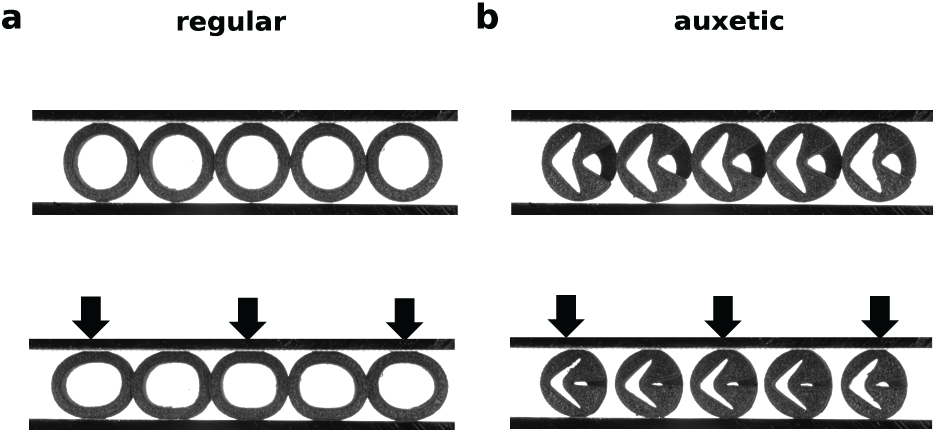}
\caption{\textbf{The Poisson effect in an experiment} \textbf{(a)} The compression of a row of regular grains. A force in the transverse direction is building up starting from the first moment of compression due to the positive Poisson's ratio of the individual grains. \textbf{(b)} The compression of a row of auxetic grains. The negative Poisson's ratio causes the grains to avoid neighbouring contact for which no force chain can emerge in the transverse direction.}
\label{SI:poissons_effect}
\end{figure}

We refer to the Poisson effect as a global phenomenon where the Poisson's ratio of the individual grains determines how stresses are being formed in the transverse direction of the compaction direction. For positive Poisson's ratio grains (Fig. \ref{SI:poissons_effect}a), we observe that the individual expansion in the transverse direction causes the grains to form increasingly strong connections with neighboring grains. However, the grains in a row of auxetic grains are observed to avoid contact with their neighboring grains. As a result, there is no force chain in the transverse direction. As discussed in the main text, this effect is one of the two major components in the collapse of figures \ref{fig:3}h and i, as well as figures \ref{fig:4}d and e.

\section{Normalized experimental quantities} \label{Normalization_experimental}

In order to compare the experiments with the numerical simulations, all reported quantities have been normalized to dimensionless numbers. In the biaxial compression experiments, the measured quantities are the displacement along the diagonal axis of the square volume, the corresponding force and the initial area from the image capture. From the displacement and the initial area, we obtain the volumetric strain. The measured force is divided by the size of the two moving sides of the confining squared area to obtain the pressure. The pressure is consecutively normalized by the average individual grain stiffness over 15 percent strain corresponding to $1.80$ N/mm for auxetic grains and $2.29$ N/mm for regular grains.

The directly measurable quantities in the shearing experiments are the rotation angle, the torque, the normal displacement, the normal force and the initial height from an image capture. 

The volumetric strain as reported in figures \ref{fig:3}f,g and \ref{fig:4}b,c is calculated directly from the initial volume and the normal displacement. The shear modulus as reported in figures \ref{fig:3}f,g,h,i is calculated from a linear fit of the torque against the rotational angle. The Lissajous figure of a typical cyclic shearing measurement is shown in figure \ref{SI:extract_shearing}a. The slope is measured both for the first part of the cycle where the torque increases along with the rotation in positive direction as well as when the rotation decreases in negative direction from the initial zero degrees deflection. The slope is averaged for both positive and negative directions for ten cycles.

In order to obtain unit-less values to compare with the numerical simulations, the slope is divided by both the radius of the cylinder as well as the circular contact. The rotational angle is converted to a rotational distance and the shear strain is next obtained by estimating the shear band to be 10 cm based on empirical observations. In an equation form:


\begin{figure}[h]
    \centering
\includegraphics[width=0.5\linewidth]{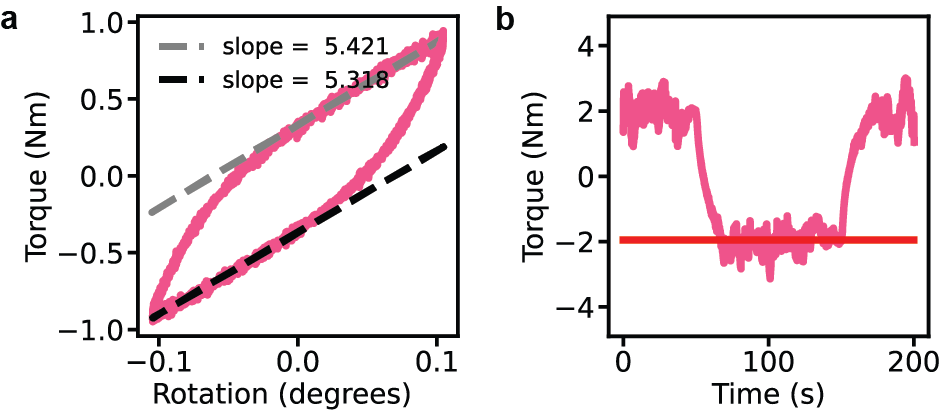}
\caption{\textbf{Extracting shear modulus and dynamic yield stress} \textbf{(a)} A typical Lissajous curve of a small strain cyclic shearing experiment. During the initial rotation from 0 to 0.1 degrees, the slope of the torque against the rotational angle is extracted with a linear fit. Once returned at the 0 degree angle and moving towards -0.1 degrees, the slope is again extracted with a linear fit. The shear modulus is calculated from averaging both shearing directions for 10 cycles. \textbf{(b)} A typical profile of the torque against the time for a large strain cyclic shearing experiment. When the imposed rotation is in the positive direction, the torque is positive. We see two transitions during a full cyclic period. The dynamic yield stress is calculated from the averaged measured torque on during the rotation from 100 degrees towards -100 degrees.}
\label{SI:extract_shearing}
\end{figure}

\begin{align*}
    G = \frac{\textrm{torque}}{\textrm{angle}}  \frac{1}{r}   \frac{1}{2\pi  r} \frac{360 \cdot  \textrm{shear band}}{2\pi  r},
\end{align*}

where $G$ is the shear modulus and $r$ the radius of the cylindrical shear cell. The shear modulus is consecutively normalized by the average individual grain stiffness over 15 percent strain corresponding to 1.80 N/mm for auxetic grains and 2.29 N/mm for regular grains.

The normal pressure as reported in figures \ref{fig:3}h,i and \ref{fig:4}d,e is calculated by dividing the measured normal force by the contacting circular area. To obtain unit-less values from this pressure as well, we normalize the pressure by the individual grain stiffness.


The yield stress as reported in figures \ref{fig:4}b,c,d,e is calculated from the average torque in a large strain shearing experiment (fig \ref{SI:extract_shearing}b). This value is divided by the circular radius and the circular contacting area after which a normalization by the corresponding individual grain stiffness gives the reported results.

\section{Numerical model for all Poisson's ratio grains} \label{Num_model}

Here we explain the proposed a numerical model for deformable bidimensional grains. We will consider a collection of elliptical bidimensional grains, where each grain has 5 degrees of freedom: 2 translational $\vec{x}_i$, 1 rotational $\theta_i$, and 2 of deformation, the semi-axis $a_i$ and $b_i$, see Fig.\ref{SI:Overlap}a. The chosen elliptical shape for the grains is arbitrary, and other shapes are applicable to this model, e.g. spherocylinders.

The elliptical grains will interact with a repulsive potential proportional to their overlap $\Delta_{ij}$, with a stiffness coefficient $k$. Each grain will have an area restitutive energy $(A_i - A_{i}^0)^2$, with a stiffness $k_A$, where $A_i=a_ib_i$ is proportional to the particle area, and $A_i^0$ is proportional to the particle's initial area. And each grain will have a shape restitutive energy $\epsilon_i^2$, where $\epsilon_i=a_i-b_i$ is a measurement of the linear eccentricity. Thus, at rest and with no stress, the grains are circles of area $\pi A_i^0$. We will use a constant $C$ to normalize the potential such that any grain has the same linear response regardless of their Poisson's ratio. Putting it all together, the potential energy for this model is:

\begin{align}
    V = C \left(\sum_{<i,j>}\frac{k}{2}\Delta_{ij}^{2} + \sum_{i}\frac{k_{A}}{2A_{i}^0}(A_i - A_{i}^0)^2 + \sum_i\frac{k_s}{2}\epsilon_i^2 \right).
\end{align}

Notice that the area energy is normalized by the initial area, to keep the correct units for the stiffness. As there is no explicit expression for the overlap between two ellipses, we use an approximation for this overlap distance, where we consider that the overlap happens in the line connecting the centers of two grains. Thus this distance is given by $\Delta_{ij}=R_i+R_j-|\vec{x}_j-\vec{x}_i|$, with $R_i=\sqrt{\left(a_i \cos{(\theta_i-\alpha_{ij})}\right)^2+\left(b_i \sin{(\theta_i-\alpha_{ij})}\right)^2}$, and $\alpha_{ij}=\arctan{\frac{y_i-y_j}{x_i-x_j}}$ is the angle of the distance vector between grains centers. More about this approximation in Appendix \ref{sec:Overlap}. We replace the stiffness constants, by the ratios $\mu=\frac{k_a+k_s}{k}$ and $\lambda=\frac{k_A}{k_s}$. Thus the system potential is:

\begin{align}
    V = C k\left(\sum_{<i,j>}\frac{1}{2}\Delta_{ij}^{2} + \sum_{i}\frac{\lambda\mu}{2A_{i}^0 (\lambda+1)}(A_i - A_{i}^0)^2 + \sum_i\frac{\mu}{2(\lambda+1)}\epsilon_i^2\right).
    \label{eq:potential}
\end{align}

Here $\mu$ controls the ratio between the particle deformation and the particle overlap energies. And $\lambda$ controls how the particle will deform.

If $\mu \ll 1$, the particle's will mostly deform, without overlapping at all. In this scenario, if we set $\lambda \ll 1$, the particles will change their shape while keeping their area constant, i.e. they will deform elliptically like regular grains with a positive Poisson's ratio. On the other hand, if we set $\lambda \gg 1$, the particles will deform while keeping their shape or eccentricity constant, i.e. the particles will shrink like auxetic grains with a negative Poisson's ratio.

\subsection{Single Particle Compression}

In this section we will discuss the linear response of a single grain under the potential of Eq.\ref{eq:potential}. We will use the linear response to fix the value of the normalization constant $C$. Lastly we will check on the relationship between the grains coefficients $\lambda$ and $\mu$, with the Poisson's ratio. 

\begin{figure}
    \centering
\includegraphics[width=1.0\linewidth]{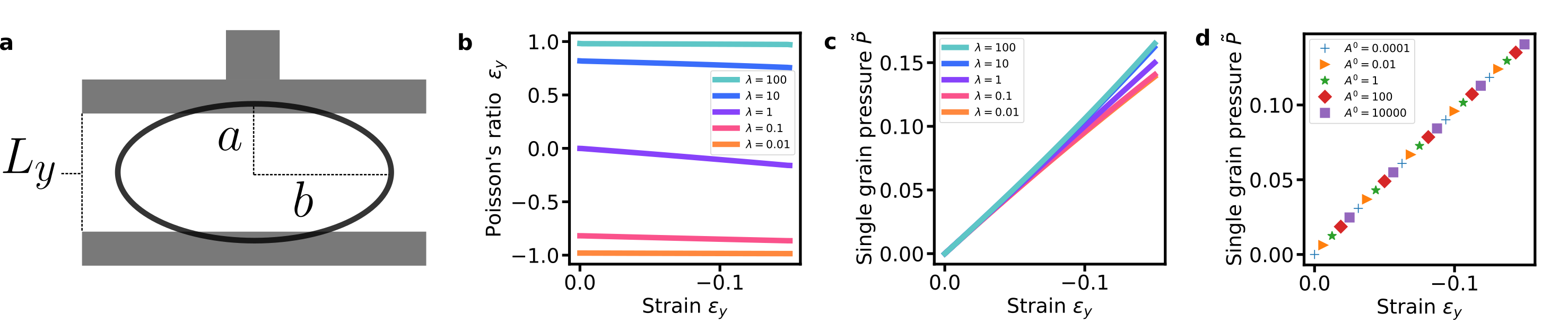}
\caption{\textbf{Single grain compression.} \textbf{(a)} Schematic of a single grain compression. \textbf{(b)} Poisson's ratio of a single grain, as a function of the strain $\epsilon_y=\frac{L_y}{\sqrt{A^0}}-1$. \textbf{(c,d)} Normalized vertical pressure $\tilde P=\frac{F_y}{k2b}$ of a single grain. In \textbf{(c)} we compare different values of $\lambda$. Notice that the grains all start with the same linear response. In \textbf{(d)} we compare the strain-stress curve of particles with different sizes $A^0$, with $\lambda=0.1$. We see that the response is independent of the initial area. $\mu=0.1$ in all of these tests.}
\label{SISP}
\end{figure}

First, we can write the potential energy for the compression of a single particle, as seen in Fig.\ref{SISP}:

\begin{align}
    V = Ck\left(\left(\frac{L_y}{2}-a\right)^{2} +\frac{\lambda\mu}{2A^0 (\lambda+1)}\left(ab - A^0\right)^2 + \frac{\mu}{2(\lambda+1)}\left(a-b\right)^2\right).
\end{align}

\noindent Here $L_y$ is the distance between two compressing plates, $a$ and $b$ are the particles semi-axes and $A^0$ is the initial area. We expand around the initial area $L_y=\delta L_y + 2\sqrt{A^0}$, $a=\delta a +\sqrt{A^0}$, $b=\delta b +\sqrt{A^0}$:

\begin{align}
    V = Ck\left(\left(\frac{\delta L_y}{2}-\delta a\right)^2 +\frac{\lambda\mu}{2 (\lambda+1)}\left(\frac{\delta a\delta b}{\sqrt{A^0}}+\delta a+\delta b\right)^2 + \frac{\mu}{2(\lambda+1)}\left(\delta a-\delta b\right)^2\right).
\end{align}

\noindent From this equation, we can measure the vertical force on the compressing piston:
\begin{align}
    F_y=-\frac{\partial V}{\partial \delta L_y}=Ck\left(\delta a-\frac{\delta L_y}{2}\right).
\end{align}
And we can measure the force that deforms the particle axes:
\begin{align}
    &\frac{\partial V}{Ck\partial \delta a}=\frac{\lambda\mu}{(\lambda+1)}(\delta a+\delta b)+\frac{\mu}{\lambda+1}(\delta a-\delta b)+ 2k\left(\delta a-\frac{\delta L_y}{2}\right),\\
    &\frac{\partial V}{Ck\partial \delta b}= \frac{\lambda\mu}{(\lambda+1)}(\delta a+\delta b)+\frac{\mu}{\lambda+1}(\delta b-\delta a).
\end{align}
As the particle is in equilibrium, both these forces are equal to zero. From where we can determine that:
\begin{align}
    F_y=-Ck\delta L_y\left(2+\frac{(\lambda+1)^2}{\lambda \mu}\right)^{-1}.
\end{align}

\noindent The linear response scales with the particle's elastic properties $\lambda$ and $\mu$. But we want all the particles to have the same linear response. So we normalize the potential energy by setting 
\begin{align}
    C=2+\frac{(\lambda+1)^2}{\lambda \mu}.
    \label{eq:normalization}
\end{align}

Now that we have the value of $C$, we simulate the compression of a single particle on 5 grains with different $\lambda$, at $\mu=0.1$. In Fig.\ref{SISP}b we see that each particle has a distinct Poisson's ratio that at linear strain behaves like $\nu=\frac{\lambda-1}{\lambda+1}$. In the plot of In Fig.\ref{SISP}c and Fig.\ref{SISP}d we confirm that with the normalization all the particles have the same linear response, regardless of it's Poisson's ratio or the particles initial size $A^0$. In all our numerical simulations we use the same values for our elastic coefficients, with $\lambda=(0.01,\ 0.1,\ 1,\ 10,\ 100)$. Which approximately corresponds to a Poisson's ratio of $\nu=(-1,\ -0.8,\ 0,\ 0.8,\ 1)$ in a linear regime.

\subsection{Particle Forces}

To simulate a system of grains, we need the force over each particle, we calculate them from the potential energy Eq.\ref{eq:potential}. There is a total of $5$ forces each one for each degree of freedom. Each axis of a particle $i$ experience a deformation force:
\begin{align}
    F_{a_i}=-\frac{dV}{da_i} = -kC\left(\sum_{j}\Delta_{ij}\frac{a_i \cos{(\theta_i -\alpha_{ij})}^2}{R_i} +\frac{\lambda\mu}{A_{i}^0 (\lambda+1)}(a_ib_i-A_{i}^0)b_i + \frac{\mu}{(\lambda+1)}(a_i-b_i)\right),
\end{align}
\begin{align}
    F_{b_i}=-\frac{dV}{db_i} = -kC\left(\sum_{j}\Delta_{ij}\frac{b_i \sin{(\theta_i -\alpha_{ij})}^2}{R_i} +\frac{\lambda\mu}{A_{i}^0 (\lambda+1)}(a_ib_i-A_{i}^0)a_i + \frac{\mu}{(\lambda+1)}(b_i-a_i)\right).
\end{align}

\noindent The sum over $j$ is a sum over the neighbour of particle $i$. 
The angle of the particle experiences a torque:
\begin{align}
    F_{\theta_i}=-\frac{dV}{d\theta_i} = -kC\sum_{j}\Delta_{ij}\left(a_i^2-b_i^2\right)\sin(\alpha_{ij}-\theta_i)\cos(\alpha_{ij}-\theta_i)
\end{align}
And the position of the particle feels a force:
\begin{align}
\label{eq:Fx}
    F_{x_i}=-\frac{dV}{dx_i} = -kC\sum_{j}\Delta_{ij}\left( \cos(\alpha_{ij}) - \frac{dR_i}{dx_i}-\frac{dR_j}{dx_i} \right),
\end{align}
\begin{align}
    F_{y_i}=-\frac{dV}{dy_i} = -kC\sum_{j}\Delta_{ij}\left( \cos(\alpha_{ij}) - \frac{dR_i}{dy_i}-\frac{dR_j}{dy_i} \right),
\end{align}
where the derivatives $\frac{dR_i}{dx_i}$ and $\frac{dR_i}{dy_i}$ are given by:
\begin{align}
    \frac{dR_i}{dx_i} &= \frac{1}{R_ir_{ij}}\left( a_i^2 \cos(\alpha_{ij}-\theta_i)\left(\cos(\theta_i)-\cos(\alpha_{ij})\cos(\alpha_{ij}-\theta_i)\right)+b_i^2 \sin(\alpha_{ij}-\theta_i)\left(-\sin(\theta_i)-\cos(\alpha_{ij})\sin(\alpha_{ij}-\theta_i)\right)\right),\\
    \frac{dR_i}{dy_i} &= \frac{1}{R_ir_{ij}}\left( a_i^2 \cos(\alpha_{ij}-\theta_i)\left(\sin(\theta_i)-\sin(\alpha_{ij})\cos(\alpha_{ij}-\theta_i)\right)+b_i^2 \sin(\alpha_{ij}-\theta_i)\left(\cos(\theta_i)-\sin(\alpha_{ij})\sin(\alpha_{ij}-\theta_i)\right)\right).
\end{align}

\subsection{Ellipse Overlap Approximation}
\label{sec:Overlap}

\begin{figure}
    \centering
\includegraphics[width=1.0\linewidth]{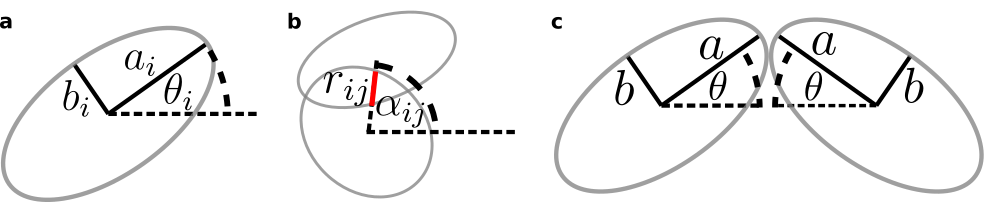}
\caption{\textbf{Overlap approximation.} \textbf{(a)} Degrees of freedom of a deformable grain. \textbf{(b)} Schematic of the overlap between two particles. In red is the overlap distance $r_{ij}$, which is measured on the segment from one particle's center to the other. The angle between particle's centers is $\alpha_{ij}$. \textbf{(c)} Two exact particles touching each other. The interaction will happen at the rightmost/leftmost section of each particle.}
\label{SI:Overlap}
\end{figure}

Even though there are several numerical algorithms to measure the overlap area between two ellipses \cite{OverlapArea}. It is impossible to have an exact explicit formula for this area. Usually an algorithm is used to find a solution. But as we want a differentiable function, we opted for an approximated approach. In this approximation we consider that the intersection between two ellipses happens in the line that connects their centers, as shown in red in Fig.\ref{SI:Overlap}b. This approximation is exact in the trivial case of two circles or two aligned ellipses. To show that a force calculated through this approximation is well behaved, in the following we will use a test case where we can exactly find the overlap between two ellipses, and we will compare it against this approximation.

Consider the test case in Fig.\ref{SI:Overlap}c where two identical ellipses interact. Both have the same orientation angle $\theta$. In the leftmost ellipse, the distance from its center to any point with angle $\alpha$ in the border of the ellipse is given by 

\begin{align}
    R=\sqrt{\left(a\cos(\theta-\alpha)\right)^2+\left(b\sin(\theta-\alpha)\right)^2}.
\end{align}
Notice that the interaction between ellipses happens at the rightmost point of this ellipse. Thus, the overlap between both ellipses is given by $d-2R(\alpha^*)\cos(\alpha^*)$. Where $d$ is the distance between the ellipse's centers, and $\alpha^*$ is the angle at the contact between both ellipses, given by extremizing the projection of the left ellipse
\begin{align}
\label{eq:extremize}
    \frac{R(\alpha)\cos(\alpha)}{d \alpha}\bigg\rvert_{\alpha=\alpha^*}=0.
\end{align}
If we consider a linear spring on this overlap, the overlap force in the test case would be
\begin{align}
    F^t_x=kC(2R(\alpha^*)\cos(\alpha^*)-d).
\end{align}
We can compare it against the force in our approximation, using Eq.\ref{eq:Fx}, where $\alpha^*=0$
\begin{align}
    F_x=kC(2R(0)-d).
\end{align}

We now perturb both of these expression around small changes in shape, given by $a=b(1+\epsilon)$ with $\epsilon \ll 1$.
First need to get the contact angle and its derivatives for $\epsilon=0$. From Eq.\ref{eq:extremize}
\begin{align}
    \alpha^*|_{\epsilon=0}&=0,\\
    \frac{d\alpha^*}{d\epsilon}\bigg\rvert_{\epsilon=0}&=\sin(2\theta).
\end{align}
Using these we now expand the force in both cases:
\begin{align}
    F^t_x=kCb\left(2-\frac{d}{b}+\epsilon \cos^2(\theta)+\epsilon^2 5\frac{1-\cos(4\theta)}{16}+O(\epsilon^3)\right),
\end{align}
\begin{align}
    F_x=kCb\left(2-\frac{d}{b}+\epsilon \cos^2(\theta)+\epsilon^2\frac{1-\cos(4\theta)}{16} +O(\epsilon^3)\right).
\end{align}
We see that $F^t_x$ is equal to $F_x$ up until the second order in $\epsilon$, where there is a difference by a factor of $5$. Thus we proved that at least in this test case the used approximation is well behaved.

\section{Numerical biaxial setup} \label{guillotine_numerical}

\begin{figure}
    \centering
\includegraphics[width=1.0\linewidth]{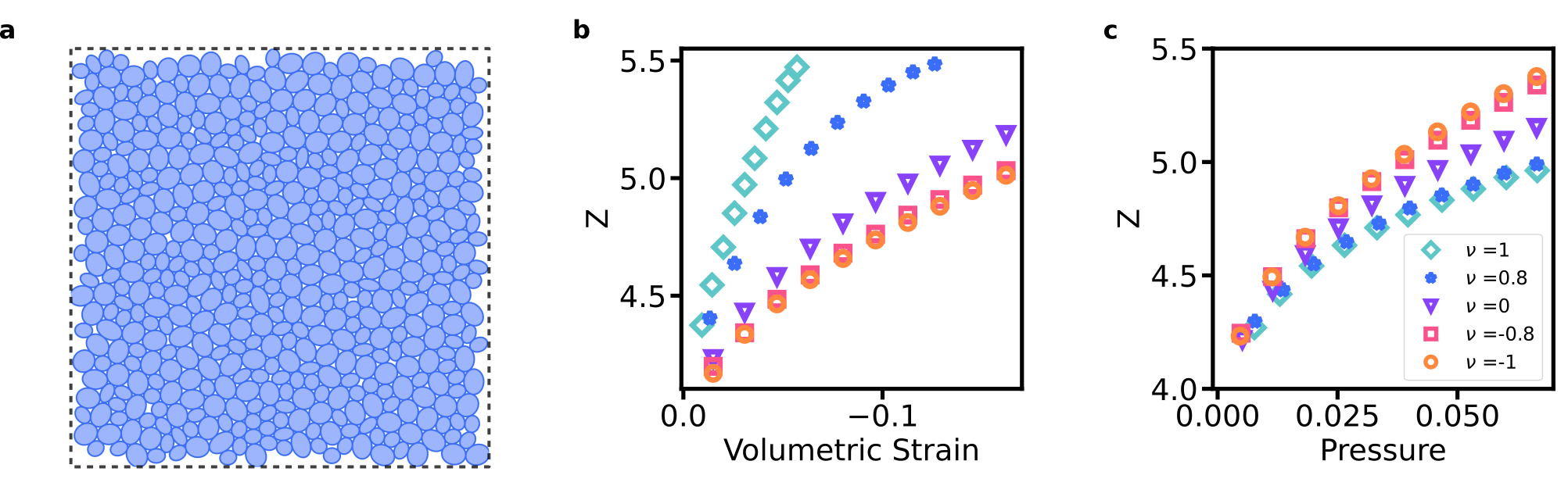}
\caption{\textbf{Coordination of deformable grains.} \textbf{(a)} Setup of the numerical biaxial experiments. The dotted line represents the periodic boundaries. \textbf{(bc)} Coordination as a function of the volumetric strain and pressure. These measurements where taken from the biaxial experiments.}
\label{SI:bulksetup}
\end{figure}

To numerically test the bulk response of each packing, we performed frictionless biaxial compression simulations. We used a system of bidisperse grains, with a radius ratio of 1.4 between big and small grains. For each Poisson's ratio, we simulated 16 bidisperse systems of 500 particles with periodic boundary conditions as seen in Fig.\ref{SI:bulksetup}. We compressed it by slowly shrinking the size of the periodic box, both directions where compressed equally. Thus the volumetric strain is $\epsilon_v=\frac{L^2}{L^2_0}-1$, where $L$ is the size of the periodic box, and $L_0$ is the size at zero pressure. At each numerical step the system's energy was relaxed using FIRE algorithm \cite{FIRE}.

As the forces in the packing are pair-wise, we measured the pressure in the periodic packing via the equation \cite{virialTheorem}:
\begin{align}
    P=\frac{1}{2 L^2}\sum_c \vec{F}^c_{\vec{x}} \cdot  \vec{x}^c,
\end{align}
where $P$ is the pressure, $\vec{F}^c_{\vec{x}}$ is the position force at contact $c$,  $\vec{x}^c$ is the distance between the pair of particles in contact $c$, the sum is over the contacts in the packing. To measure the stress in other directions, the following equation can be used

\begin{align}
    \sigma_{ij}=\frac{1}{L^2}\sum_c {F}_{x_i}^c x_{j}^c.
    \label{eq:stress}
\end{align}

\section{Numerical shearing setup} \label{shear_setup_numerical}

\begin{figure}
    \centering
\includegraphics[width=1.0\linewidth]{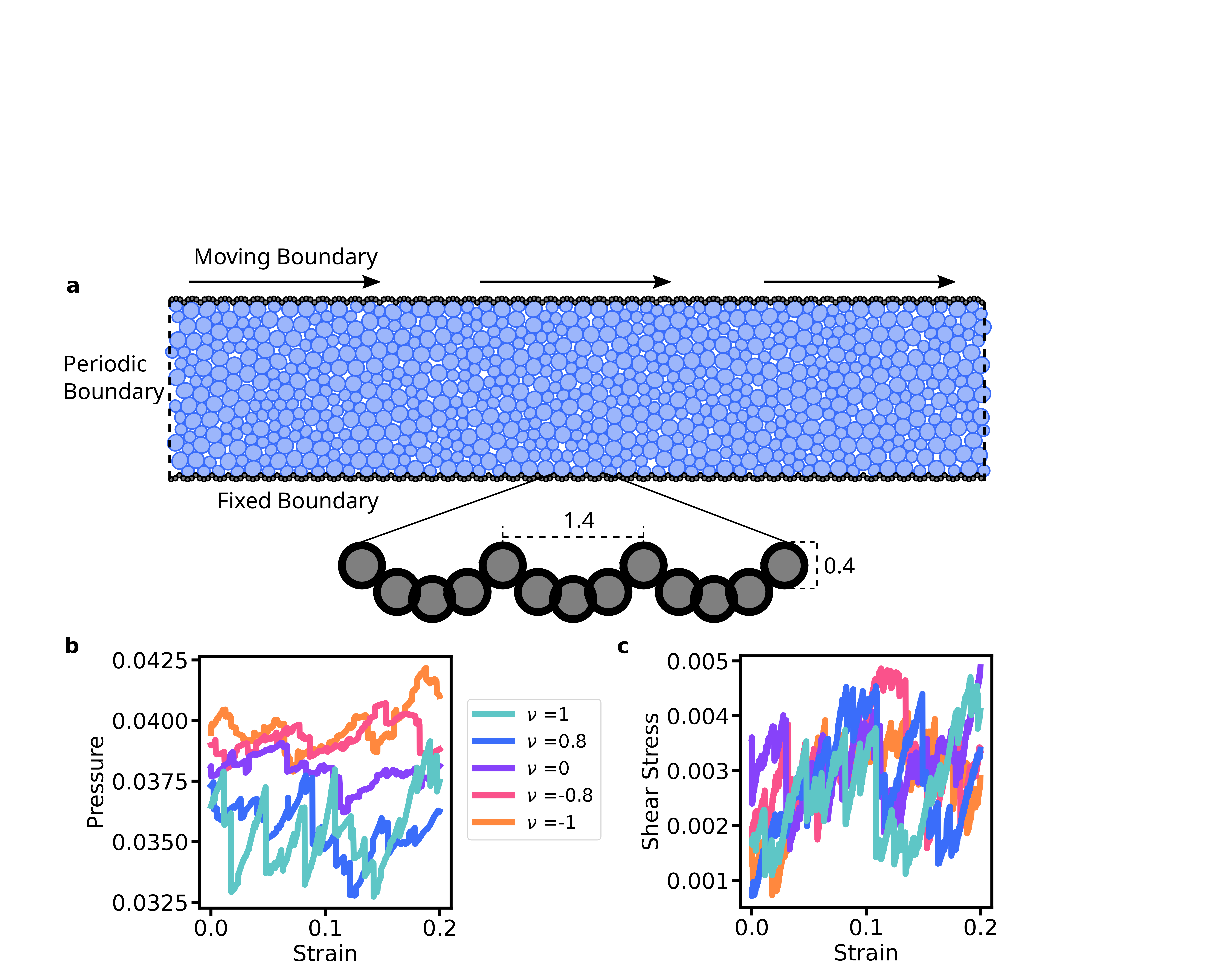}
\caption{\textbf{Numerical Couette Cell.} \textbf{(a)} Example setup for the shearing simulations. In blue are the system's grains. The dotted line represents a periodic boundary in the x axis. The grey particles create two boundaries at the top (moving) and bottom (fixed) of the system. The boundary's teeth are such that they can hold a particle of radius $0.7$. \textbf{(bc)} Samples of the pressure/shear stress against shear strain for this experiment. We see that the pressure has smaller fluctuations for the auxetic grains, meanwhile, the shear stress has no significant change as a function of the Poisson's ratio.}
\label{SI:shearsetup}
\end{figure}

In the shearing numerical simulations we emulated the experimental setup, with the main difference that we neglected friction in the simulations. We used bidisperse packings of $800$ particles, with a ratio of $1.4$ between big and small particles. The system has fixed boundaries in the vertical axis, and is periodic on the horizontal axis. The fixed boundaries are composed by multiple rigid particles arranged in a serrated pattern, imitating the experimental boundaries as seen in Fig.\ref{SI:shearsetup}. The boundary is formed by fixed soft particles which are $3.5$ times smaller than the big particles, these particles have the same forces as any other particle, but they can't deform. Together they form concave semi-circular shapes where a single big grain can exactly fit. The horizontal size of the system is fixed, and its length is $100$ times the radius of a big particle.
Stress on the fixed boundaries can be measured either by using Eq.\ref{eq:stress}, or by adding the forces applied to the boundaries.

As in the experimental setup (See Appendix \ref{shear_setup_experimental}), we performed shear strain test on a Couette cell system. While shearing, the system's volume remained fixed, and the vertical load was allowed to fluctuate. At each time step a strain rate of $10^{-4}$ was applied on the top boundary, and the system's energy was relaxed using the FIRE algorithm \cite{FIRE}.

In Fig.\ref{SI:shearsetup}b and c, we see samples of the pressure and shear stress as a function of the shear strain.  We observe that the pressure fluctuations are much smaller for auxetic systems than non-auxetic. This is mainly reflected in Fig.\ref{fig:5} of the main text, where we show that contact force fluctuations are smaller for auxetic systems. On the other hand, the shear stress doesn't show a big difference as a function of auxeticity. The shear modulus in main text Fig.\ref{fig:3} is measured from the derivative of the shear stress. Meanwhile, the yield stress is measured from the average shear stress, after it reaches a steady state.

\section{Numerical contact number and Poisson effect} \label{poisson_effect_numerical}

In Fig.\ref{SI:bulksetup}b we observe that the coordination $Z$ (the average number of contacts per particle) measured at a fixed volume of the system, is lower for auxetic particles. This fits the known result that $Z$ is higher for elliptical particles compared with disks \cite{van2009jamming}. Surprisingly, when we do the same measurements at fixed pressure Fig.\ref{SI:bulksetup}c, $Z$ is higher for auxetic particles.

\section{Normalized numerical quantities} \label{Normalization_numerical}

The most important normalization performed, was setting $C$ in Eq.\ref{eq:normalization}. Such that all grains have the same elastic response. Other than this, all pressures/stress are normalized by the single particle stiffness. Such that they can be compared against the experiments.s

\end{appendix}

\end{document}